%% file: galaxies_fermi_arXiv.tex
\documentclass{emulateapj}

\usepackage{apjfonts}
\usepackage{natbib,graphicx,epsfig,rotating,color,verbatim,amsmath,booktabs}

\shorttitle{GeV Observations of Star-forming Galaxies with \textit{Fermi} LAT}
\shortauthors{The \textit{Fermi} LAT Collaboration}

\def\mathbi#1{\textbf{\em #1}}

\global\let\tablenotetext\relax

\begin{document}

\title{GeV Observations of Star-forming Galaxies with \textit{Fermi} LAT}

\input{author_list.tex}

\keywords{cosmic rays --- Galaxies: starburst ---
Gamma rays: galaxies --- Gamma rays: diffuse background}

\begin{abstract}
 
Recent detections of the starburst galaxies M82 and NGC 253 by
gamma-ray telescopes suggest that galaxies rapidly forming massive
stars are more luminous at gamma-ray energies compared to their quiescent relatives. 
Building upon those results, we examine a sample
of 69 dwarf, spiral, and luminous and ultraluminous
infrared galaxies at photon energies 0.1--100 GeV using 3 years of
data collected by the Large Area Telescope (LAT)
on the \textit{Fermi Gamma-ray Space Telescope} (\textit{Fermi}). Measured
fluxes from significantly detected sources and flux
upper limits for the remaining galaxies are used to
explore the physics of cosmic rays in galaxies. We find further
evidence for quasi-linear scaling relations between gamma-ray
luminosity and both radio continuum luminosity and total infrared
luminosity which apply both to quiescent galaxies of the
Local Group and low-redshift starburst galaxies (conservative $P$-values $\lesssim0.05$ accounting for statistical and systematic
uncertainties). The normalizations of these scaling relations
correspond to luminosity ratios of $\log(L_{0.1-100 \; \rm{GeV}}/L_{1.4
\; \rm{GHz}}) = 1.7 \pm 0.1_{\rm (statistical)} \pm 0.2_{\rm (dispersion)}$
and $\log(L_{0.1-100 \; \rm{GeV}}/L_{8-1000 \; \mu\rm{m}}) = -4.3 \pm 0.1_{\rm
(statistical)} \pm 0.2_{\rm (dispersion)}$ for a galaxy with a star
formation rate of 1 $M_{\odot}$ yr$^{-1}$, assuming a Chabrier initial mass function. Using the relationship
between infrared luminosity and gamma-ray luminosity, the collective
intensity of unresolved star-forming galaxies at redshifts
$0<z<2.5$ above 0.1 GeV is estimated to be 0.4--2.4 $\times 10^{-6}$ ph cm$^{-2}$
s$^{-1}$ sr$^{-1}$ (4--23\% of the intensity of the isotropic diffuse
component measured with the LAT). We anticipate that $\sim10$ galaxies could be detected by
their cosmic-ray induced gamma-ray emission during a 10-year \textit{Fermi} mission. 

\end{abstract}

\input{sec_introduction.tex}

\input{sec_candidate_selection.tex}

\input{sec_observations_and_analysis.tex}

\input{sec_results.tex}

\input{sec_discussion.tex} 

\input{sec_conclusions.tex}

\acknowledgments

The \textit{Fermi} LAT Collaboration acknowledges generous ongoing support
from a number of agencies and institutes that have supported both the
development and the operation of the LAT as well as scientific data analysis.
These include the National Aeronautics and Space Administration (NASA) and the
Department of Energy in the United States, the Commissariat \`a l'Energie
Atomique and the Centre National de la Recherche Scientifique / Institut
National de
Physique Nucl\'eaire et de Physique des Particules in France, the Agenzia
Spaziale
Italiana and the Istituto Nazionale di Fisica Nucleare in Italy, the Ministry of
Education, Culture, Sports, Science and Technology (MEXT), High Energy
Accelerator Research
Organization (KEK) and Japan Aerospace Exploration Agency (JAXA) in Japan, and
the K.~A.~Wallenberg Foundation, the Swedish Research Council and the
Swedish National Space Board in Sweden.

Additional support for science analysis during the operations phase is
gratefully acknowledged from the Istituto Nazionale di Astrofisica in Italy
and the Centre National d'\'Etudes Spatiales in France. 
The Spanish Ministry of Science and the Argentinian CONICET additionally supported this work. 

We gratefully acknowledge comments from the anonymous referee which
helped clarify the scientific interpretation of the results.

K.B. thanks Eric Feigelson of the Penn State Center for
Astrostatistics for suggestions regarding the analysis of
partially censored datasets. K.B. is supported by a Stanford Graduate Fellowship.

This research has made use of NASA's Astrophysics Data System, and the
NASA/IPAC Extragalactic Database (NED) which is operated by the Jet
Propulsion Laboratory, California Institute of Technology, under contract with NASA. 

\begin{appendix}


\input{app_sec_kendall_tau.tex}

\end{appendix}

\clearpage

\bibliography{galaxies}

\clearpage

\input{table_candidates.tex}

\clearpage

\input{table_local_group.tex}

\clearpage

\input{table_results.tex}

\clearpage

\begin{figure*}[t]
\center
\includegraphics[width=.45\textwidth]{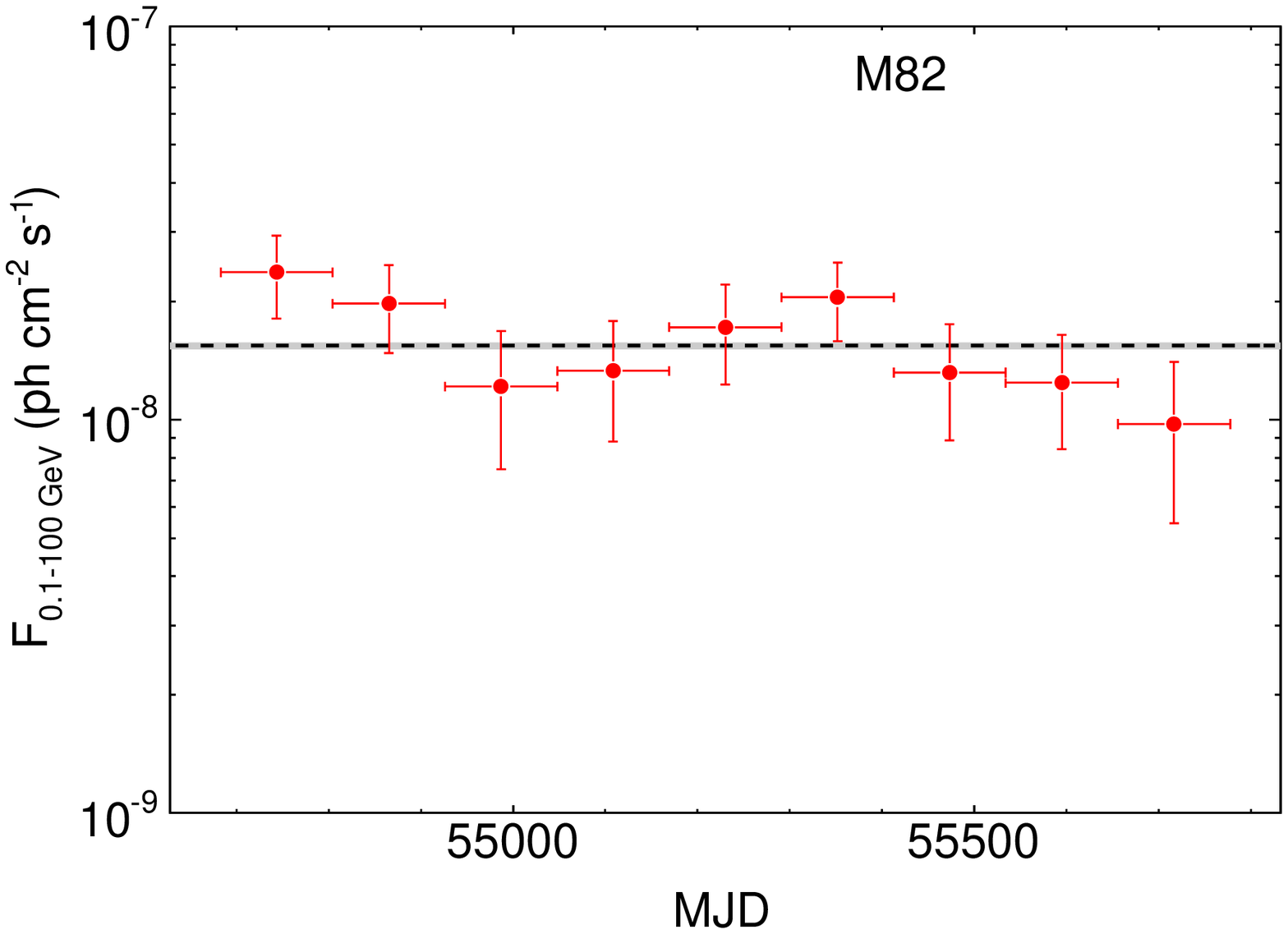}
\includegraphics[width=.45\textwidth]{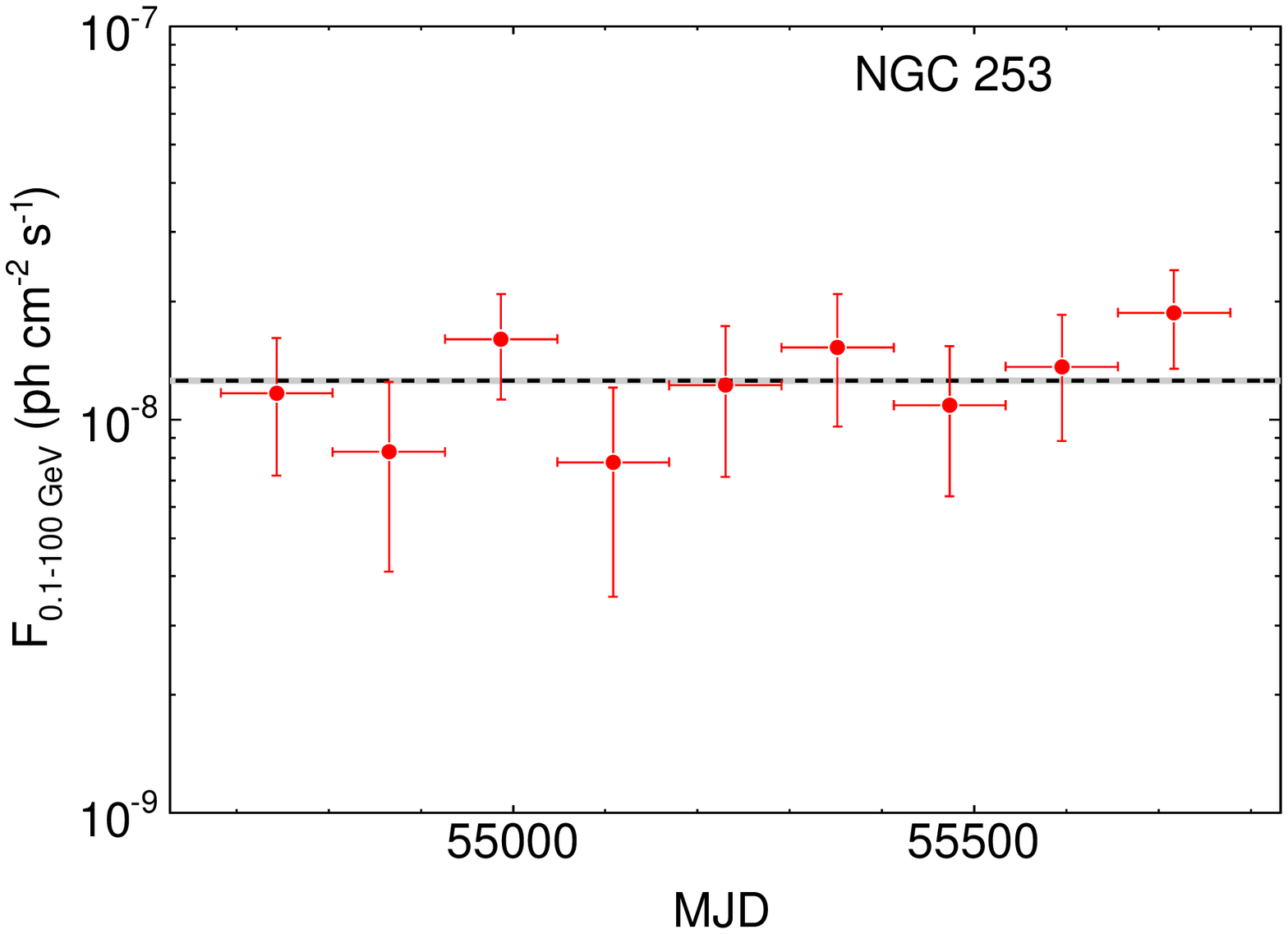}
\includegraphics[width=.45\textwidth]{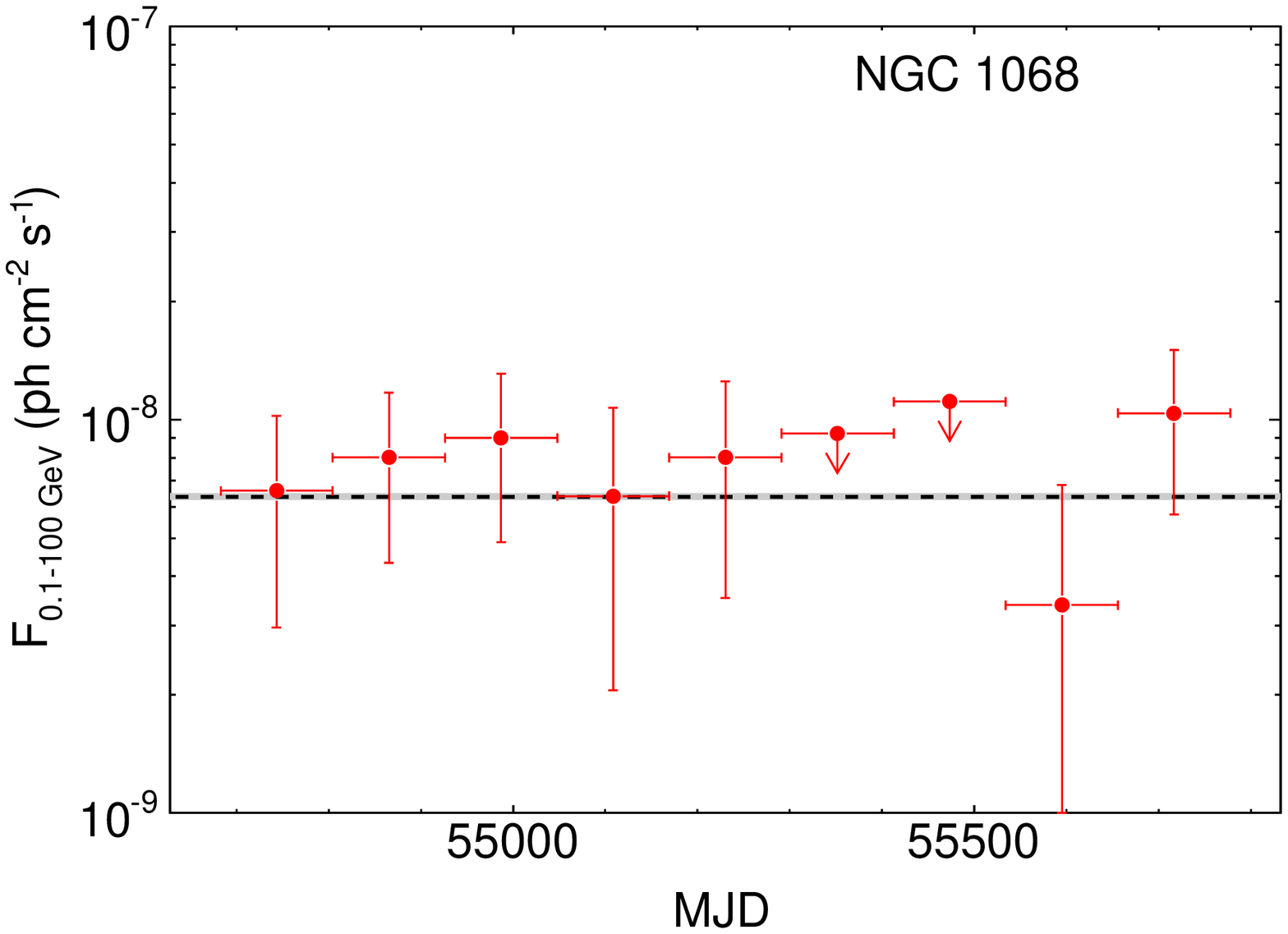}
\includegraphics[width=.45\textwidth]{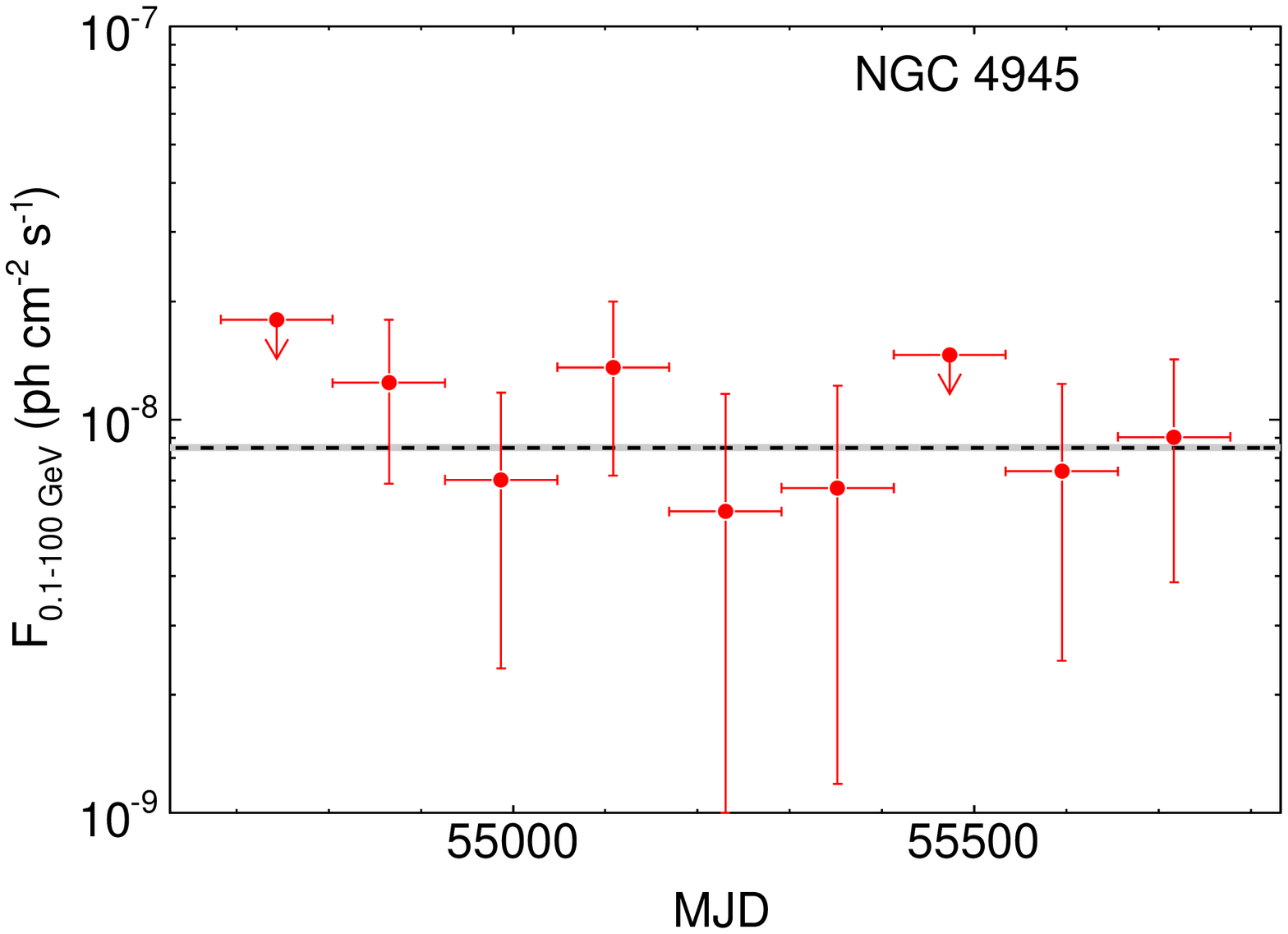}
\caption{Gamma-ray light curves for four significantly detected galaxies. The
full three-year observation period was divided into 12 time intervals of
$\sim90$ days each. A maximum likelihood fit was performed for each of the
shorter time intervals to test for variability. The dashed black line shows the maximum likelihood flux level obtained for
the full three-year observation period. The gray band represents a 2\%
systematic uncertainty in source exposure resulting from small
inaccuracies in the dependence of the instrument response on
the source viewing angle, coupled with changes in the observing
profile as the orbit of the spacecraft precesses. Flux upper limits
are shown at the 95\% confidence level.}
\vskip0.2in
\label{fig_lightcurve}
\end{figure*}

\begin{figure*}[t]
\center
\includegraphics[width=.45\textwidth]{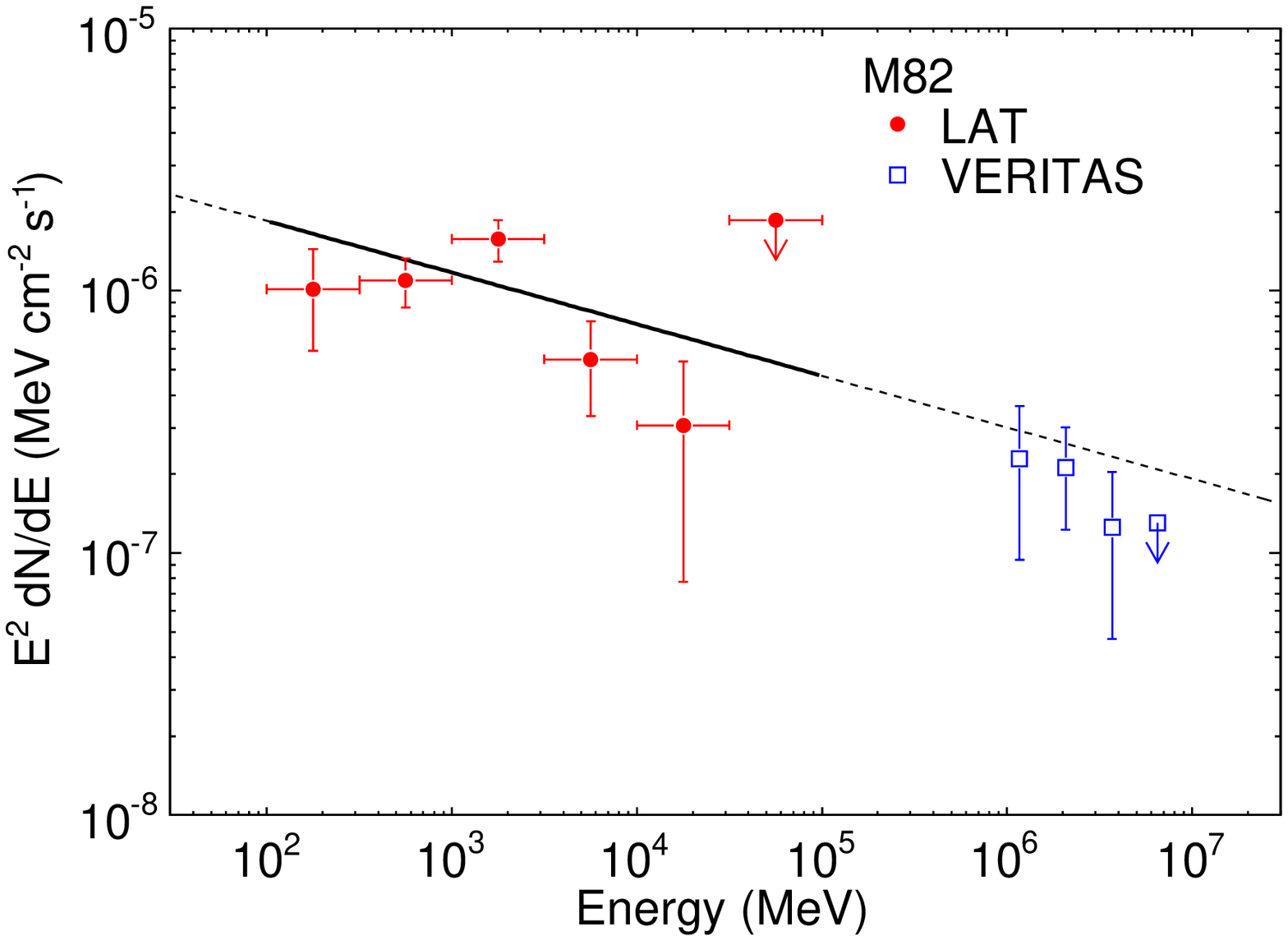}
\includegraphics[width=.45\textwidth]{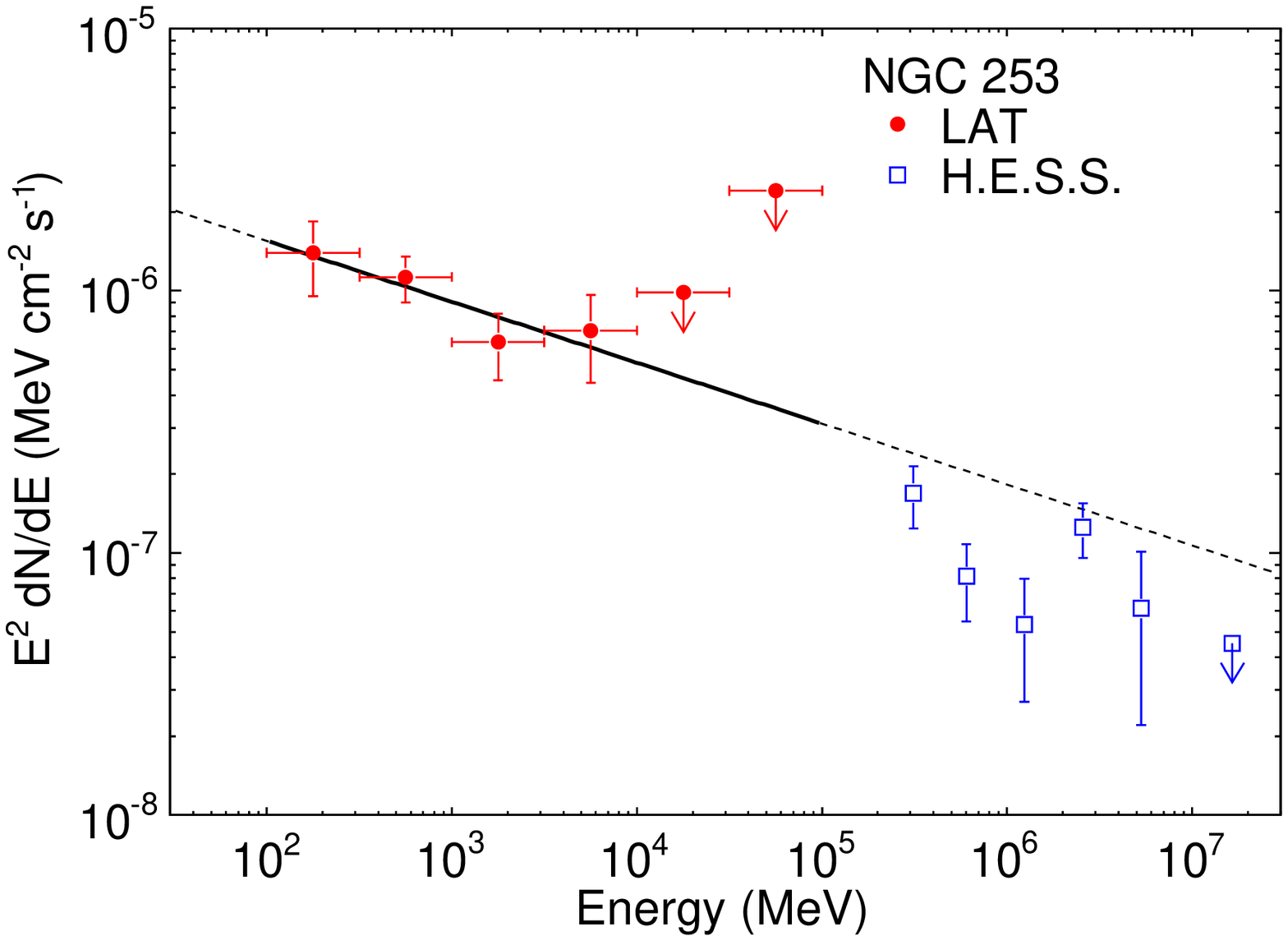}
\includegraphics[width=.45\textwidth]{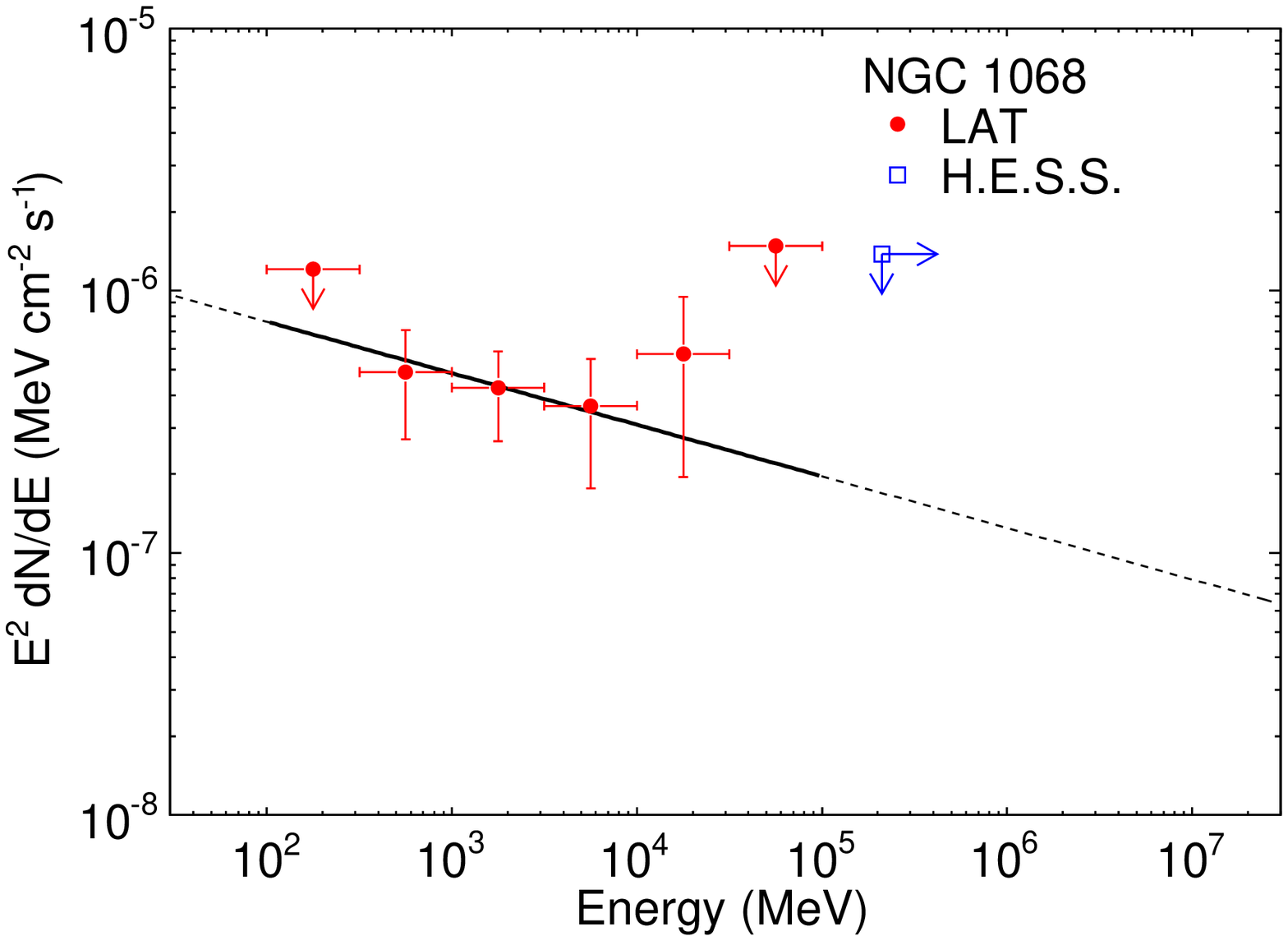}
\includegraphics[width=.45\textwidth]{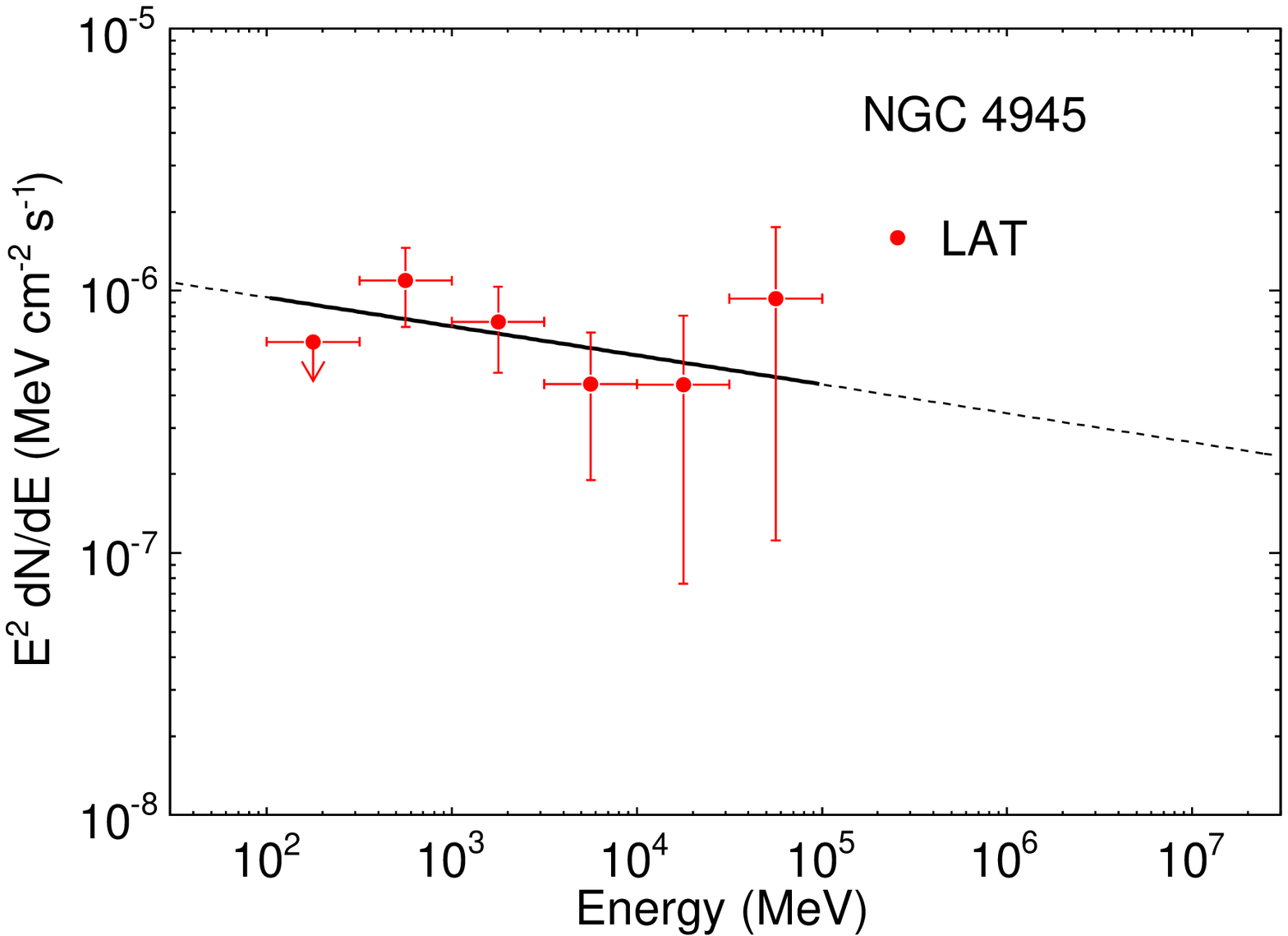}
\caption{Spectral energy distributions for four starburst galaxies
significantly detected in high energy gamma rays. 95\% confidence upper limits are
indicated in energy bins with detection significance $TS<1$. The best-fit power law
spectral model is shown for each galaxy in the energy range used for
the maximum likelihood analysis (0.1--100 GeV; solid) and in extrapolation to neighboring
wavebands (dashed). Flux measurements produced by imaging air-Cherenkov telescopes are included
for M82 \citep[VERITAS,][]{acciari_2009_m82} and NGC 253
\citep[H.E.S.S.,][]{ohm_2011_ngc_253}, and flux upper limits above 210
GeV are shown for NGC 1068 \citep[H.E.S.S.,][]{aharonian_2005_hess_ngc_1068}. The uncertainties depicted for LAT flux measurements represent
combined statistical and systematic uncertainties.}
\vskip0.2in
\label{fig_sed}
\end{figure*}

\input{table_significance.tex}

\input{table_regression.tex}



\begin{figure}[t]
\center
\includegraphics[width=0.6\textwidth]{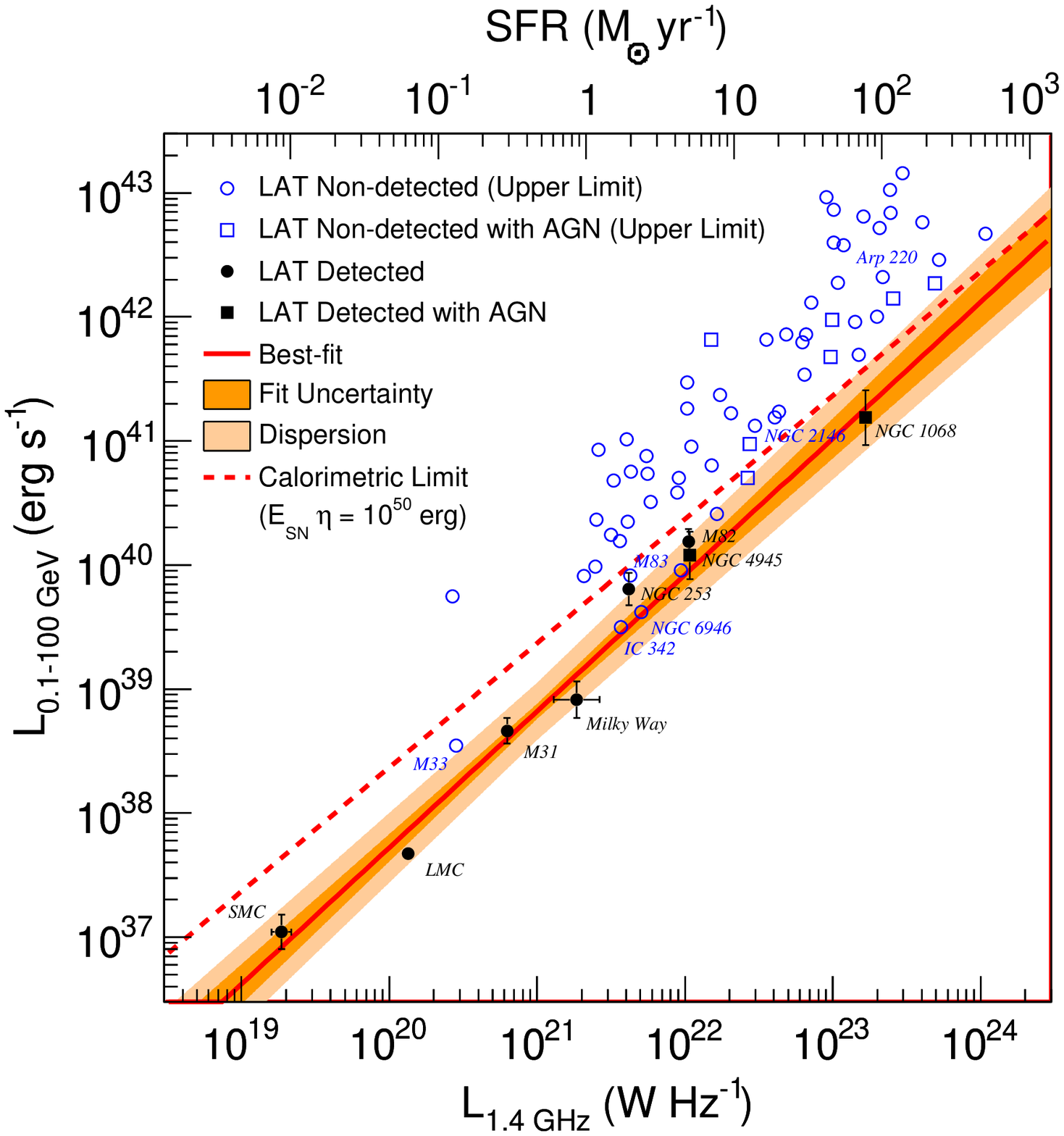}
\includegraphics[width=0.6\textwidth]{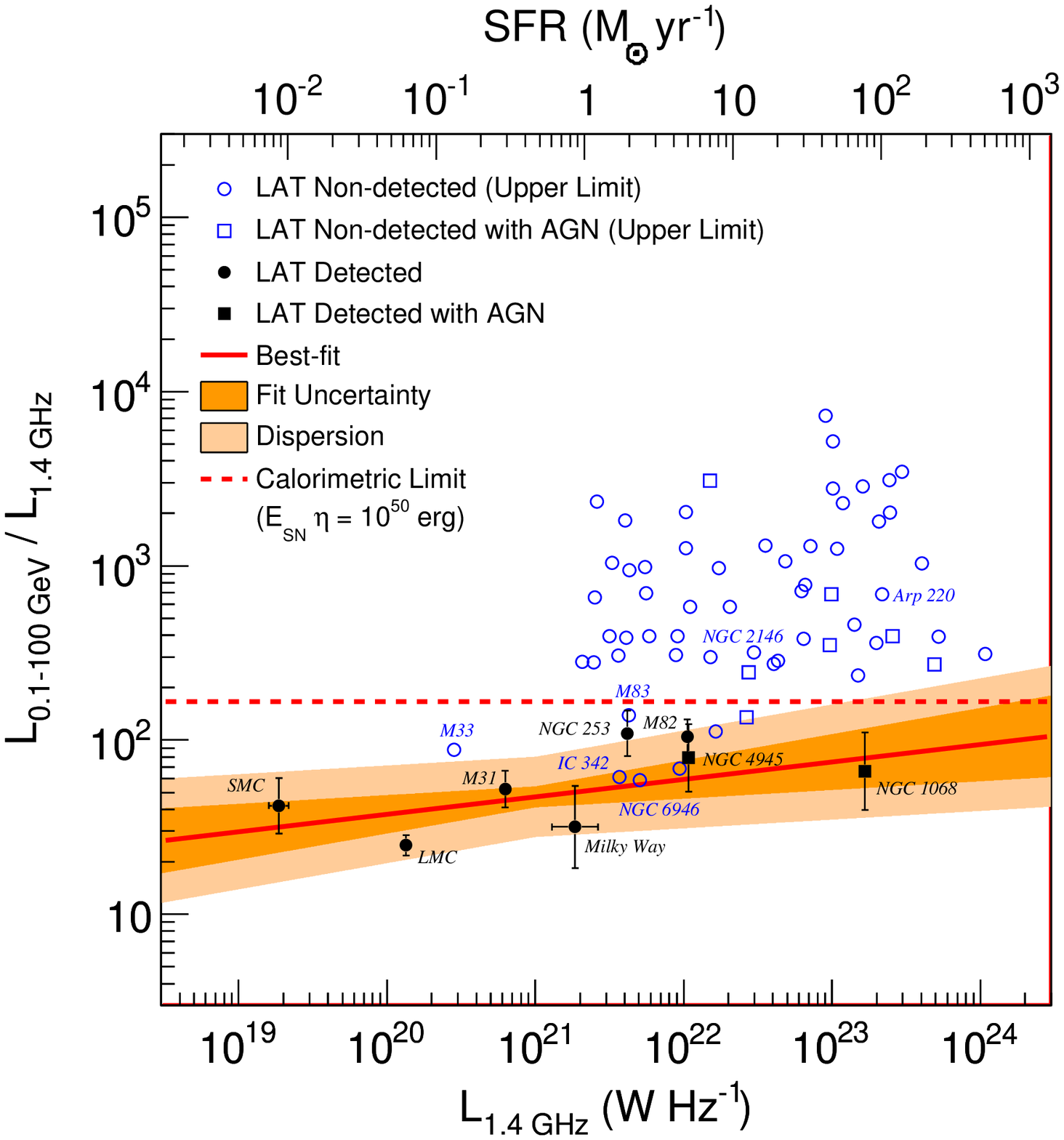}
\caption{\textit{Top Panel}: Gamma-ray luminosity (0.1--100 GeV) versus RC luminosity at 1.4 GHz.
Galaxies significantly detected by the LAT are indicated with filled
symbols whereas galaxies with gamma-ray flux upper limits (95\% confidence level) are marked with open symbols. Galaxies
hosting \textit{Swift}-BAT AGN are shown with square markers. RC luminosity
uncertainties for the non-detected galaxies are omitted for clarity, but are
typically less than 5\% at a fixed distance. The upper abscissa
indicates SFR estimated from the RC luminosity according to
equation \ref{eq_sfr_rc} \citep{yun_2001_radio}. The best-fit power law
relation obtained using the EM algorithm is shown by the red solid
line along with the fit uncertainty (darker shaded region), and intrinsic dispersion around
the fitted relation (lighter shaded region). The dashed red line
represents the expected gamma-ray luminosity in the calorimetric limit assuming an average CR
luminosity per supernova of $E_{\rm SN}\;\eta=10^{50}$ erg (see Section
\ref{subsec_cr}). \textit{Bottom Panel}: Ratio of gamma-ray luminosity (0.1--100 GeV) to RC luminosity at 1.4 GHz.}
\label{fig_rc_correlation}
\end{figure}

\begin{figure}[t]
\center
\includegraphics[width=.6\textwidth]{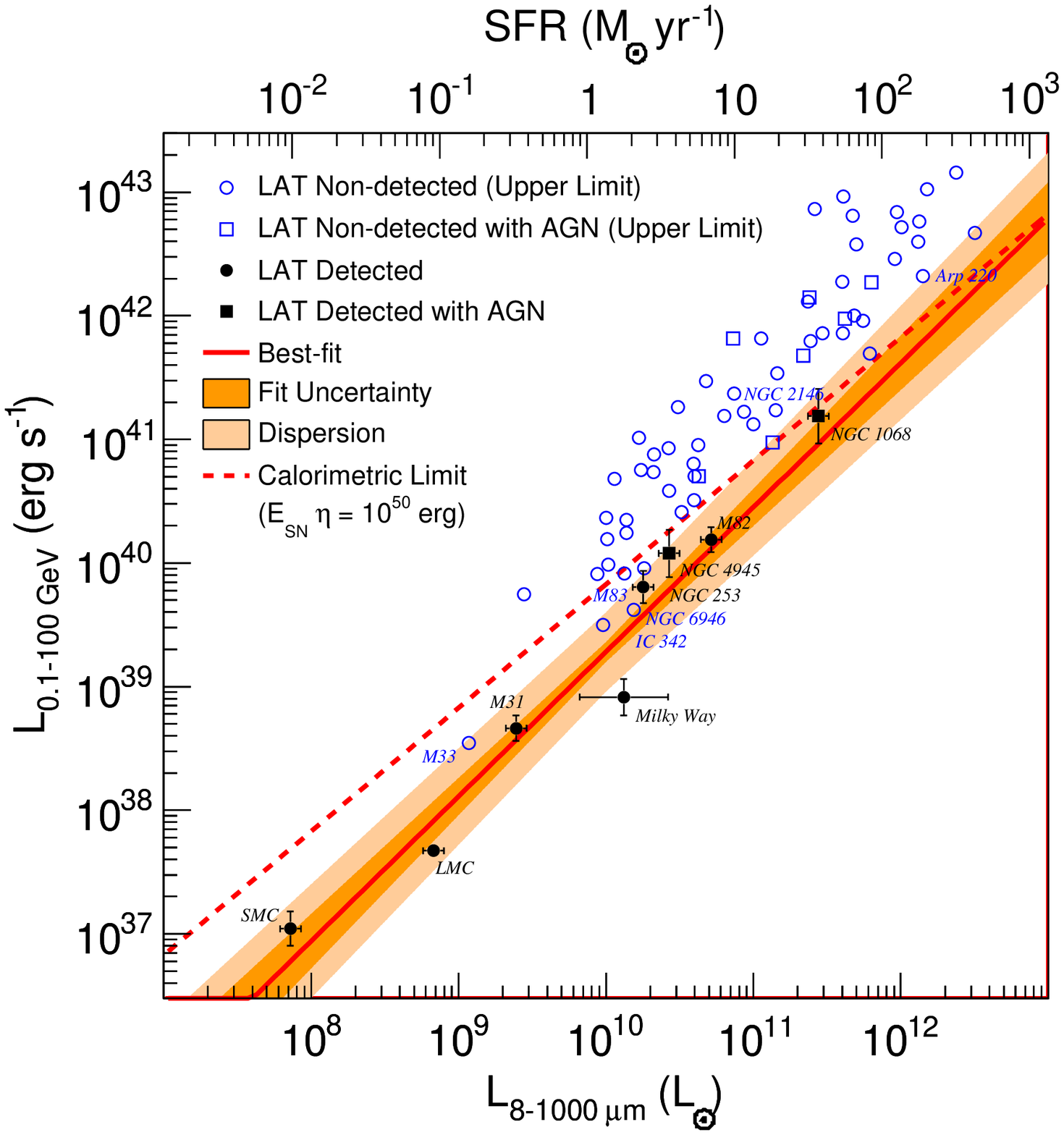}
\includegraphics[width=.6\textwidth]{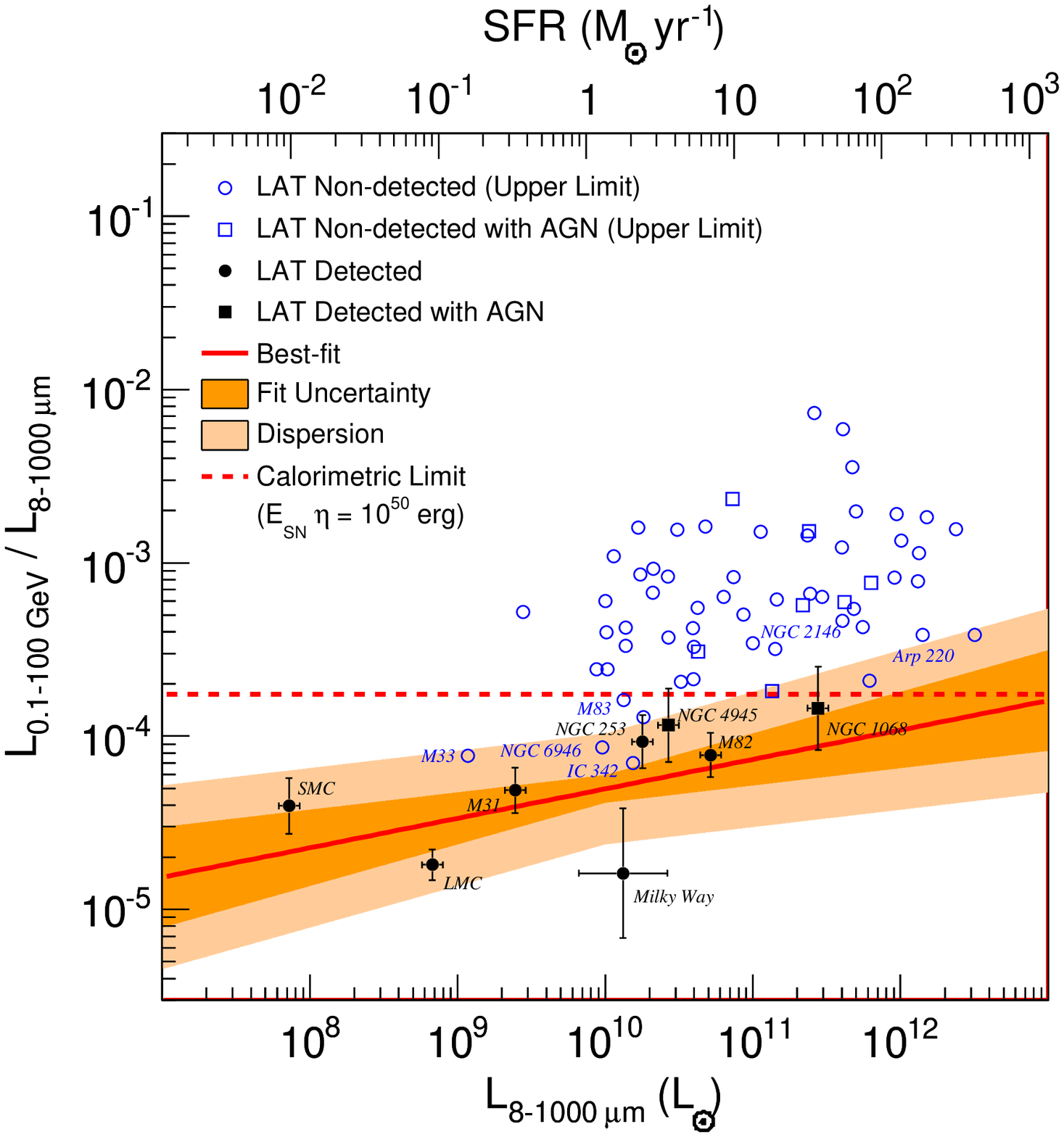}
\caption{As Figure \ref{fig_rc_correlation}, but showing gamma-ray luminosity (0.1--100 GeV) versus total IR
luminosity (8--1000 $\mu$m). IR luminosity uncertainties for the non-detected galaxies are omitted for clarity, but are
typically $\sim0.06$ dex. The upper
abscissa indicates SFR estimated from the IR luminosity according to
equation \ref{eq_sfr_ir} \citep{kennicutt_1998_schmidt}.} 
\label{fig_ir_correlation}
\end{figure}

\begin{figure}[t]
\center
\includegraphics[width=0.8\textwidth]{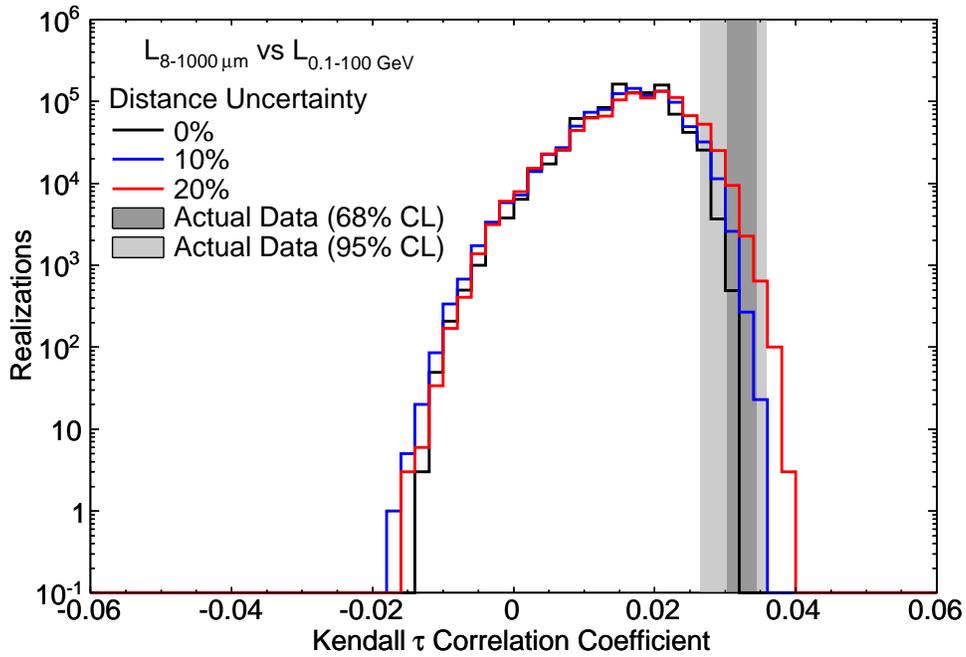}
\caption{Graphical representation of the method adopted to estimate significances of
multiwavelenth correlations using the Kendall $\tau$ statistic. Total
IR luminosity (8--1000 $\mu$m) and
gamma-ray luminosity (0.1--100 GeV) are compared using the full sample
of 69 galaxies in this example. Null hypothesis distributions of
correlation coefficients assuming independence between
wavebands are shown for 3 levels of uncertainty in distance
measurements (1 standard deviation of the scatter). The null hypothesis distributions are
generated from $10^6$ permutations of gamma-ray luminosities among
the galaxies, requiring that the resultant gamma-ray fluxes could have been
measureable given the flux sensitivity threshold of the LAT. The
correlation efficient of the actual data is represented as a
probability density to account for uncertainty in measured
gamma-ray fluxes. The gray bands indicate the inner 68\% and 95\% of
the probability density for the actual data. The correlation significance
is estimated by computing the fraction of null hypothesis realizations with correlation
coefficient larger than that obtained for the actual data (i.e. the
$P$-value). See Section \ref{subsec_multiwavelength} and Appendix \ref{app_sec_kendall_tau} for details.}
\vskip0.2in
\label{fig_kendall}
\end{figure}

\begin{figure}[t]
\center
\includegraphics[width=0.8\textwidth]{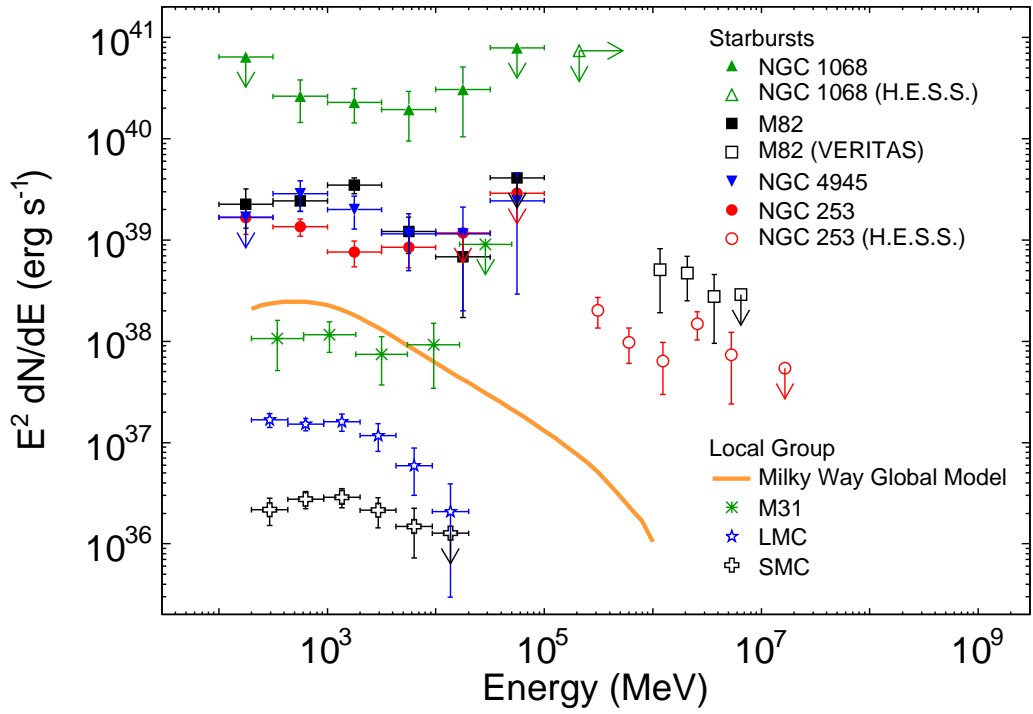}
\caption{Gamma-ray luminosity spectra of star-forming galaxies
detected by the LAT and imaging air-Cherenkov telescopes.}
\vskip0.2in
\label{fig_sed_all}
\end{figure}

\begin{figure}[t]
\center
\includegraphics[width=0.8\textwidth]{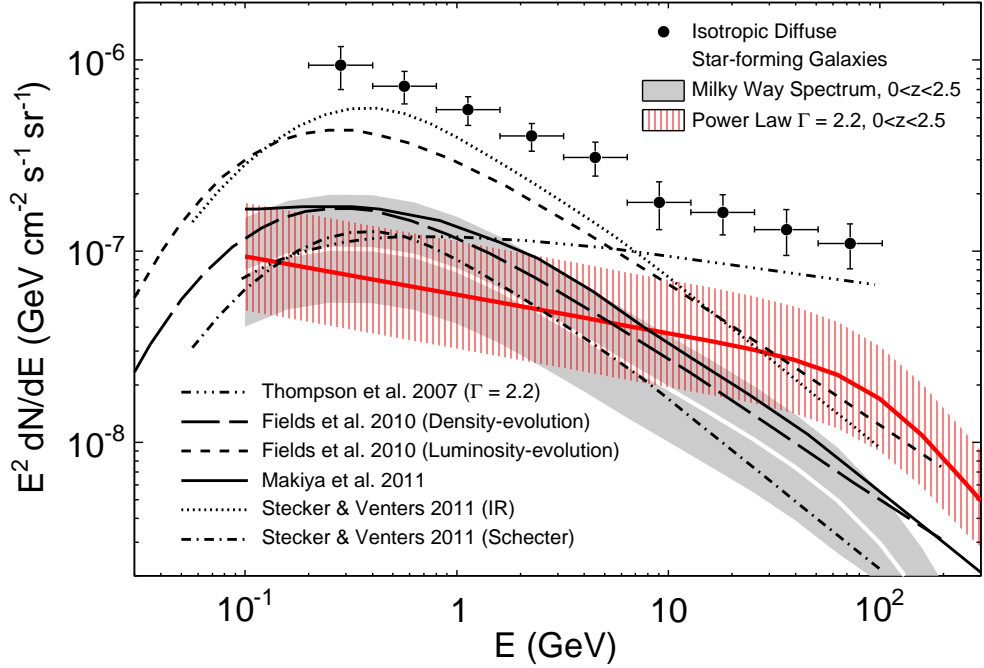}
\caption{Estimated contribution of unresolved star-forming galaxies (both quiescent and
starburst) to the isotropic diffuse gamma-ray emission measured by the
\textit{Fermi} LAT \citep[black points,][]{lat_2010_egb}. The shaded
regions indicate combined statistical and systematic uncertainties in the contributions of the respective populations. 
Two different spectral models are used to estimate the GeV gamma-ray emission from star-forming
galaxies: a power law with photon index 2.2, and a spectral
shape based on a numerical model of the global gamma-ray emission of
the Milky Way \citep{strong_2010_global}. These two spectral models should be viewed as bracketing the
expected contribution since multiple star-forming galaxy types
contribute, e.g. dwarfs, quiescent spirals, and starbursts. We consider
only the contribution of star-forming galaxies in the redshift range
$0<z<2.5$. The gamma-ray opacity of the Universe is treated using the
extragalactic background light model of \cite{franceschini_2008_ebl}. 
Several previous estimates for the intensity of unresolved
star-forming galaxies are shown for
comparison. \cite{thompson_2007_calorimeter} treated starburst
galaxies as calorimeters of CR nuclei. The normalization of the plotted curve depends on the
assumed acceleation efficiency of SNRs (0.03 in this case). The
estimates of \cite{fields_2010_galaxies_egb} and by
\cite{makiya_2011_galaxies_egb} incorporate results from the first
year of LAT observations. \cite{fields_2010_galaxies_egb} considered the extreme cases of either pure
luminosity evolution and pure density evolution of star-forming
galaxies. Two recent predictions from \cite{stecker_2011_egb_components}
are plotted: one assuming a scaling relation between IR-luminosity and
gamma-ray luminosity, and one using a redshift-evolving Schecter model to relate
galaxy gas mass to stellar mass.}
\vskip0.2in
\label{fig_egb_contribution}
\end{figure}

\clearpage

\input{table_egb_intensity.tex}

\clearpage

\begin{figure}[t]
\center
\includegraphics[width=.75\textwidth]{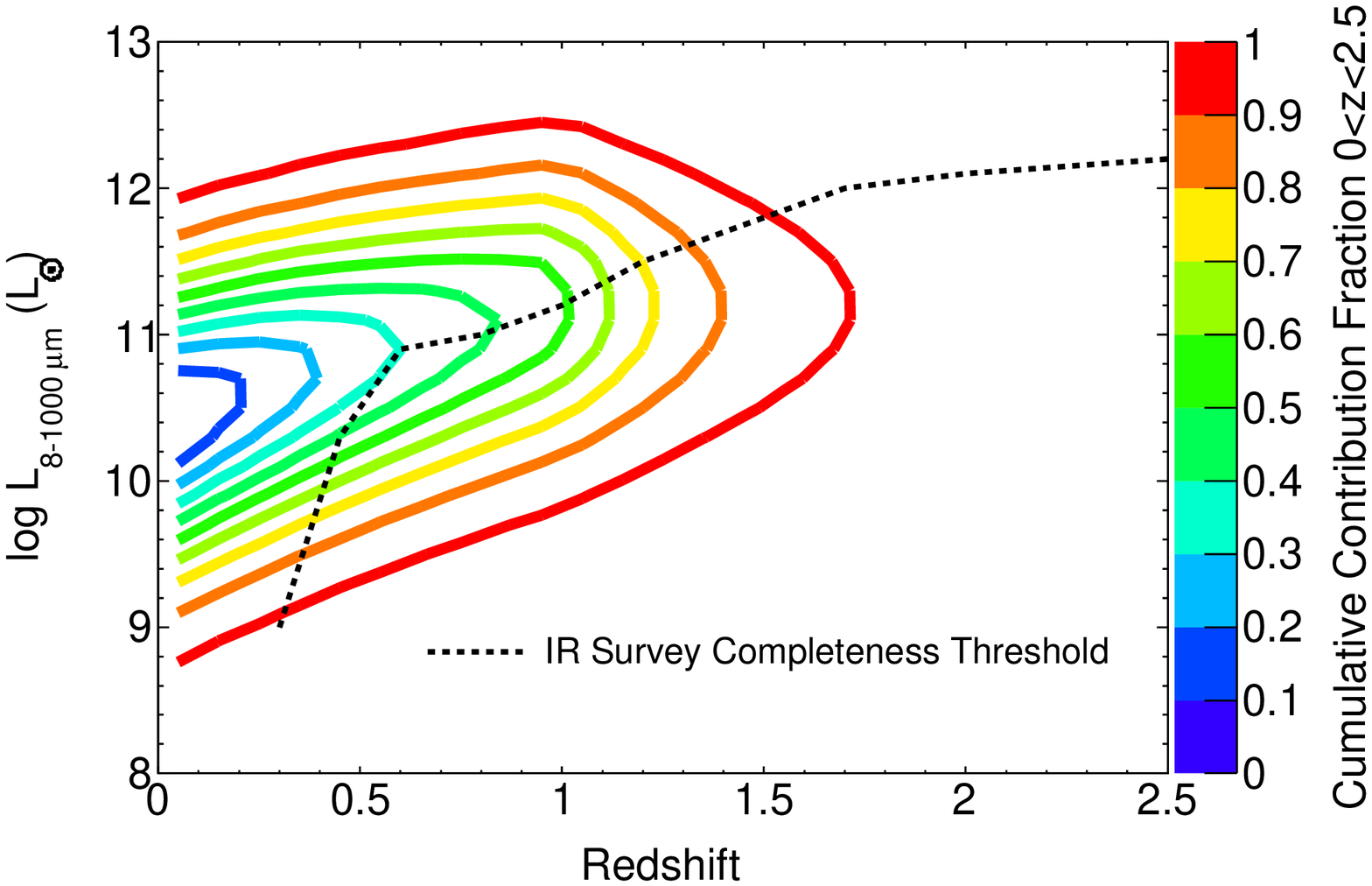}
\includegraphics[width=.75\textwidth]{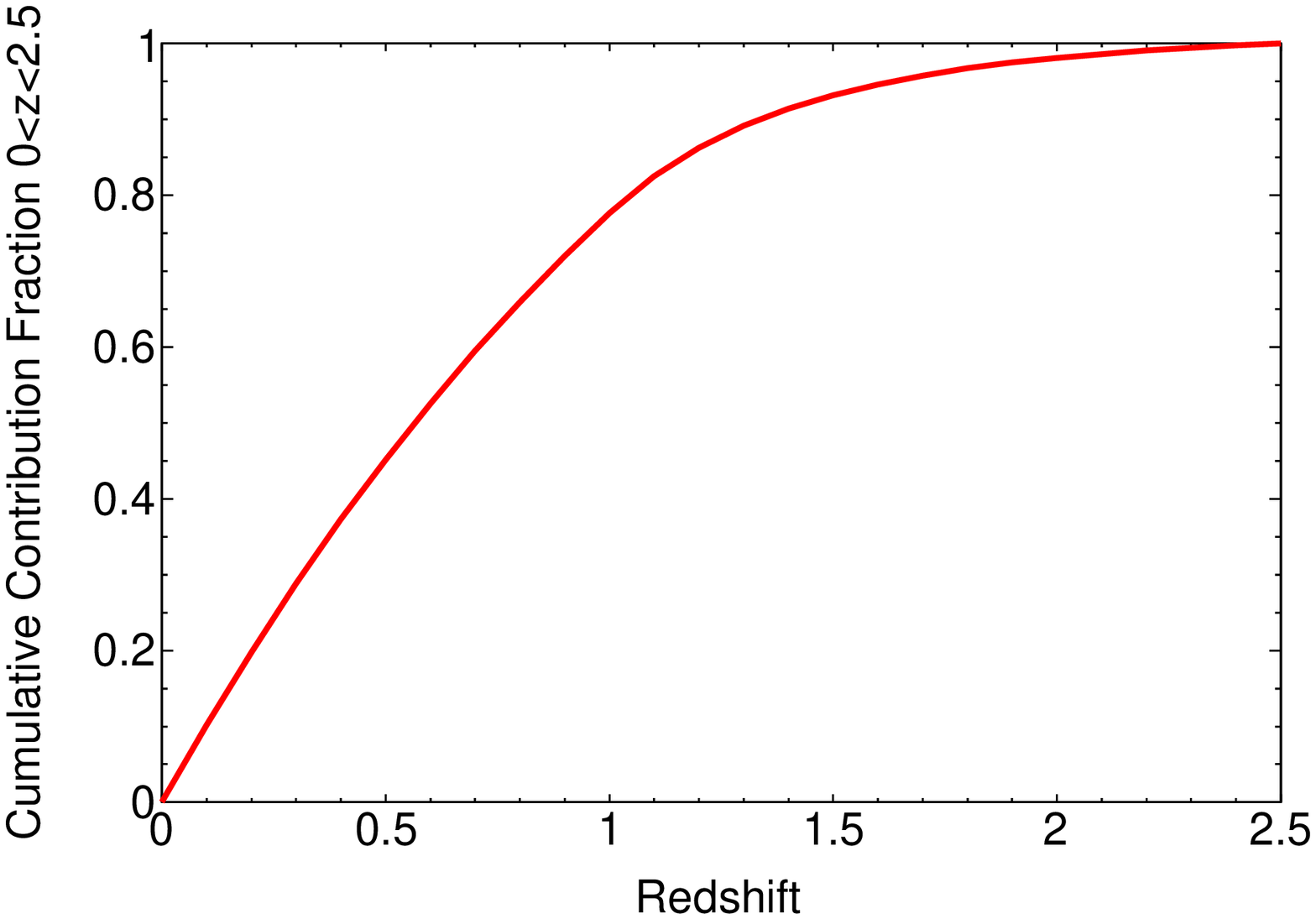}
\caption{Relative contribution of star-forming galaxies to the
isotropic diffuse gamma-ray background according to their redshift and total
IR luminosity (8--1000 $\mu$m) normalized to the total contribution in
the redshift range 0<z<2.5. \textit{Top Panel}: Solid contours indicate regions of phase
space which contribute an increasing fraction of the total energy
intensity (GeV cm$^{-2}$ s$^{-1}$ sr$^{-1}$) from all star-forming
galaxies with redshifts $0 < z < 2.5$ and $10^{8} L_{\odot} <
L_{8-1000 \; \mu{\rm m}} < 10^{13} L_{\odot}$. Contour levels are placed at 10\% intervals.
The largest contribution comes from low-redshift Milky Way
analogues ($L_{8-1000 \; \mu{\rm m}} \sim10^{10} L_{\odot}$) and
starburst galaxies comparable to M82, NGC 253, and NGC 4945. The black
dashed curve indicates the IR luminosity
above which the survey used to generate the adopted IR luminosity function is
believed to be complete \citep{rodighiero_2010_spitzer}.
\textit{Bottom Panel}: Cumulative contribution versus redshift. As above,
only the redshift range $0<z<2.5$ is considered.}
\vskip0.2in
\label{fig_egb_fractional_contribution}
\end{figure}

\begin{figure}[t]
\center
\includegraphics[width=0.8\textwidth]{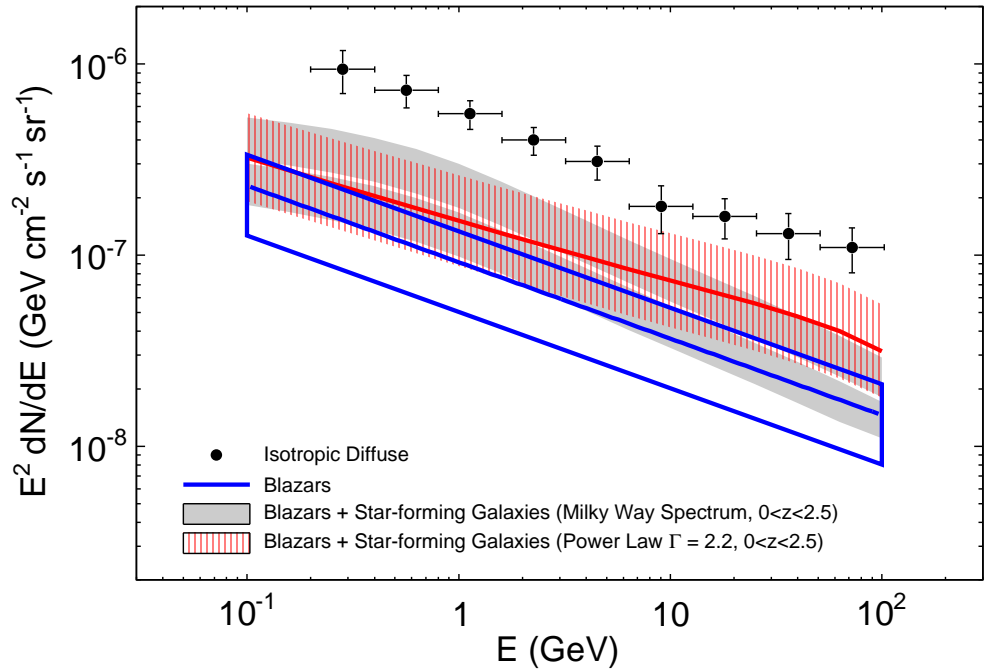}
\caption{As Figure \ref{fig_egb_contribution}, but showing the summed
contributions of blazars and star-forming galaxies (this work) to the
isotropic diffuse gamma-ray background. Two
different assumed spectral models for the star-forming galaxies are
shown. The estimated contribution of blazars is derived from the distribution of observed fluxes for high Galactic
latitude sources observed by the LAT, which are believed to
be dominated by FSRQs and BL Lac objects \citep{lat_2010_high_latitude}.}
\vskip0.2in
\label{fig_egb_contribution_sum}
\end{figure}

\clearpage

\begin{figure}[t]
\center
\includegraphics[width=0.8\textwidth]{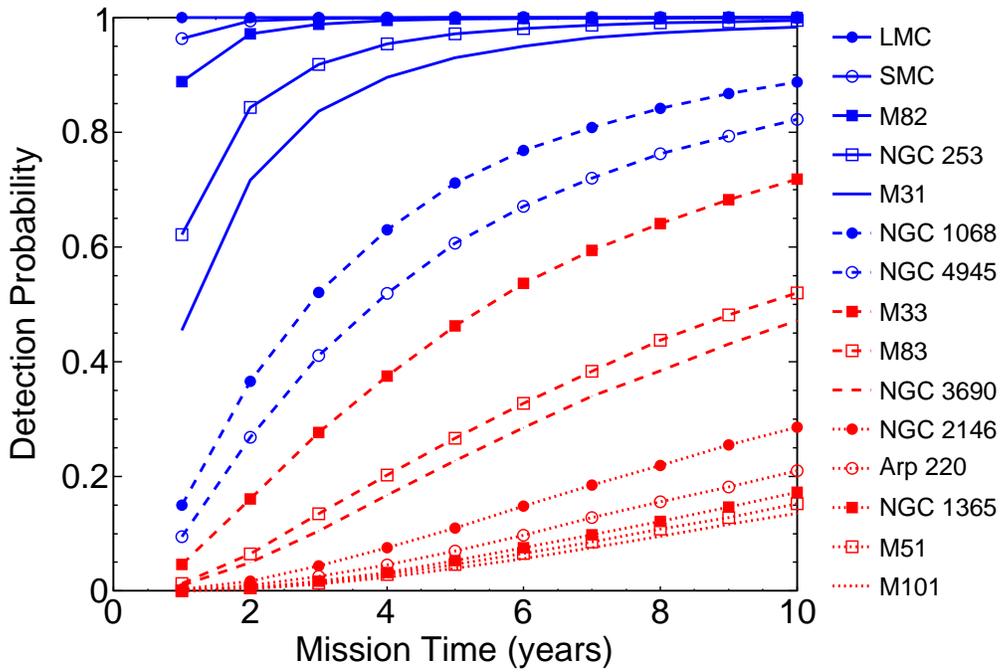}
\caption{LAT detection probabilities for individual galaxies using the
scaling relation found in Section \ref{subsec_multiwavelength}. Galaxies already detected by the LAT are
shown by blue curves, while other galaxies are drawn with red curves.}
\vskip0.2in
\label{fig_detection_prob_src}
\end{figure}

\begin{figure}[t]
\center
\includegraphics[width=0.8\textwidth]{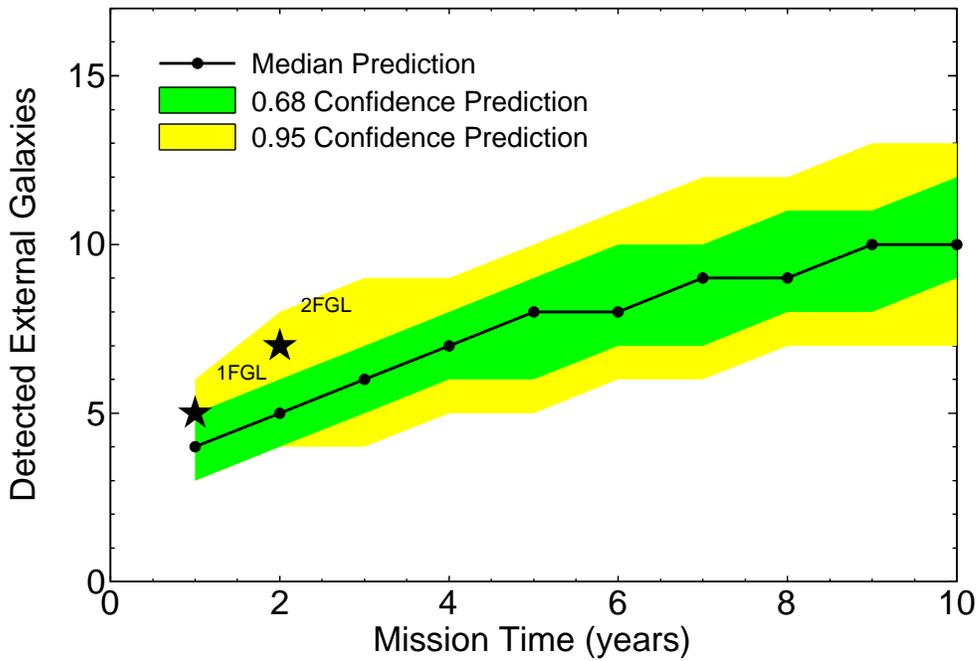}
\caption{Total number of star-forming galaxies anticipated to be
detected during a 10-year \textit{Fermi} mission. The actual numbers
of external galaxies reported in the 1FGL \citep{lat_2010_1fgl} and
2FGL \citep{lat_2012_2fgl} catalogs are marked with star symbols.}
\vskip0.2in
\label{fig_detection_prob_total}
\end{figure}

\begin{figure}[t]
\center
\includegraphics[width=.75\textwidth]{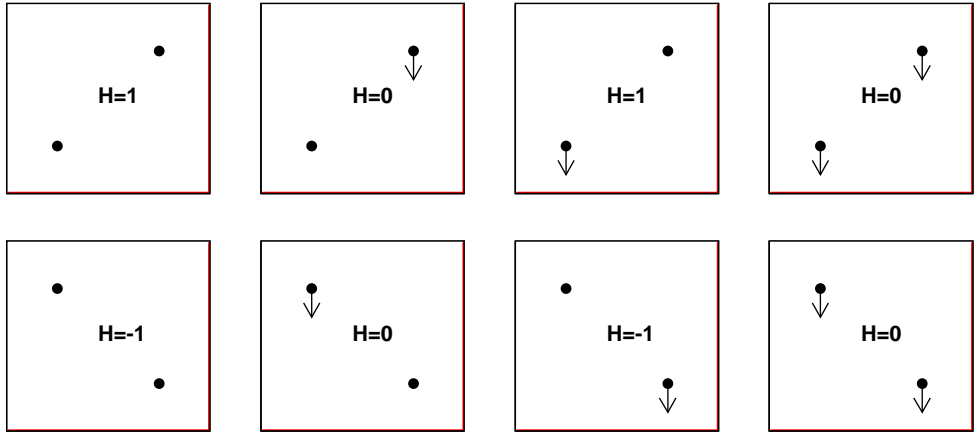}
\caption{Graphical representation of the generalized Kendall $\tau$ rank
correlation test. The $\tau$ correlation statistic is proportional
to the sum of $H$-values obtained for each pair of points in the
dataset. See Appendix \ref{app_sec_kendall_tau} for a complete desciption of
the method.}
\vskip0.2in
\label{fig_kendall_demo}
\end{figure}

\input{table_volume_limited.tex}

\input{table_flux_limited.tex}

\end{document}

%% file: author_list.tex
\author{
M.~Ackermann\altaffilmark{1}, 
M.~Ajello\altaffilmark{2}, 
A.~Allafort\altaffilmark{2}, 
L.~Baldini\altaffilmark{3}, 
J.~Ballet\altaffilmark{4}, 
D.~Bastieri\altaffilmark{5,6}, 
K.~Bechtol\altaffilmark{2,7}, 
R.~Bellazzini\altaffilmark{3}, 
B.~Berenji\altaffilmark{2}, 
E.~D.~Bloom\altaffilmark{2}
% ...
}
\altaffiltext{1}{Deutsches Elektronen Synchrotron DESY, D-15738 Zeuthen, Germany}
\altaffiltext{2}{W. W. Hansen Experimental Physics Laboratory, Kavli Institute for Particle Astrophysics and Cosmology, Department of Physics and SLAC National Accelerator Laboratory, Stanford University, Stanford, CA 94305, USA}
\altaffiltext{3}{Istituto Nazionale di Fisica Nucleare, Sezione di Pisa, I-56127 Pisa, Italy}
\altaffiltext{4}{Laboratoire AIM, CEA-IRFU/CNRS/Universit\'e Paris Diderot, Service d'Astrophysique, CEA Saclay, 91191 Gif sur Yvette, France}
\altaffiltext{5}{Istituto Nazionale di Fisica Nucleare, Sezione di Padova, I-35131 Padova, Italy}
\altaffiltext{6}{Dipartimento di Fisica ``G. Galilei", Universit\`a di Padova, I-35131 Padova, Italy}
\altaffiltext{7}{email: bechtol@stanford.edu}
% ...

%% file: sec_introduction.tex
\section{Introduction}\label{sec_introduction}


Global emission of most galaxies across much of the electromagnetic spectrum from
radio to gamma-ray energies is related to the formation and
destruction of massive stars. The exceptions are certain galaxies hosting active galactic nuclei
(AGN). Throughout this work, we refer to galaxies in which non-thermal
emission is mainly diffuse in origin rather than powered by a
supermassive black hole as `star-forming galaxies.'

O and B stars radiate the majority of their bolometric luminosity at
ultraviolet wavelengths, which is efficiently absorbed by
interstellar dust and re-emitted in the infrared
\citep[IR,][]{scoville_1983_m51}. The IR emission dominates the
spectral energy distribution of actively
star-forming galaxies \citep[reviewed by][]{sanders_1996_lirgs} and is
a robust indicator of star-formation rates \citep[e.g.][]{kennicutt_1998_hubble}. Starburst galaxies are
distinguished by kpc-scale regions of intense star-forming
activity and are often identified by their enormous IR luminosities, many being classified as luminous
infrared galaxies (LIRGs, $L_{8-1000 \; \mu{\rm m}}
> 10^{11} L_{\odot}$) and ultra-luminous infrared galaxies (ULIRGs,
$L_{8-1000 \; \mu{\rm m}}
> 10^{12} L_{\odot}$) using the definition proposed by \cite{sanders_1996_lirgs}.

Massive stars responsible for the IR emission end their lives as core-collapse
supernovae, whose remnants are believed to be the main
cosmic-ray (CR, including all species) accelerators on galactic
scales. The energy budget of CRs observed in the solar neighborhood is dominated by relativistic protons.
If $\sim10\%$ of the mechanical energy of the outgoing shocks can be transferred into CR
acceleration, supernova remnants (SNRs) would be capable of
sustaining the locally measured CR flux \citep{ginzburg_1964_origin_of_crs}.  

In star-forming galaxies, CR electrons and positrons (hereafter simply electrons) spiraling in
interstellar magnetic fields radiate diffuse synchrotron emission in
the radio continuum (RC) at a level closely related to the total IR
luminosity of the galaxies \citep{van_der_kruit_1971_radio_ir,van_der_kruit_1973_radio_ir,de_jong_1985_radio_ir,helou_1985_radio_ir,condon_1992_normal_galaxies}.
Thermal bremsstrahlung emission originating from H~{\sc II} regions ionized by the massive
stars contributes to the RC emission to a lesser extent. Remarkably,
the nearly linear empirical RC-IR correlation spans over 5 orders of
magnitude in luminosity \citep{condon_1991_radio_ir}.

CRs also create diffuse gamma-ray emission. The same CR electrons responsible for radio
synchroton emission can produce high energy radiation either through interactions with gas (bremsstrahlung) or
interstellar radiation fields (inverse Compton scattering). Inelastic collisions between CR nuclei and ambient gas
lead to the production of gamma-rays through $\pi^{0}$ decay, and also to the production of secondary CR
leptons by $\pi^{\pm}$ decay. It is therefore natural to look for correlations of gamma rays with the
CR-induced emissions at lower frequencies (and accordingly, IR emissions).


Two of the nearest starburst galaxies, M82 and NGC 253, have been
detected in high-energy gamma rays by both space-based \citep{lat_2010_starburst} and
imaging air-Cherenkov telescopes \citep{acciari_2009_m82,acero_2009_ngc_253}. Although all galaxies are expected to produce CR-induced emission at
some level, large numbers of SNRs together with dense
interstellar gas (average number densities $\sim$500 cm$^{-3}$) and intense radiation fields led several authors to
anticipate the central starbursts of M82 and NGC 253 as detectable
gamma-ray sources
\citep[e.g.][]{paglione_1996_ngc_253,blom_1999_starburst,domingo-santamaria_2005_ngc_253,persic_2008_m82,de_cea_del_pozo_2009_m82,rephaeli_2010_ngc_253}.
Because the massive stars
($M\gtrsim8 M_{\odot}$) which ultimately result in core-collapse
supernovae have lifetimes of $\mathcal{O}$(10$^{7}$ yrs), and particle
acceleration in the vicinity of a supernova happens for a short time
$\mathcal{O}$(10$^{4}$ yrs), the number of CR accelerators in a given
galaxy is thought to be closely related to the
contemporaneous star-formation rate (SFR). Indeed, the observed gamma-ray
fluxes from M82 and NGC 253 imply enhanced CR energy densitities within
galaxies undergoing rapid massive star-formation. 

A study of the Local Group galaxies detected at GeV energies, including the
Milky Way (MW), Small \citep[SMC,][]{lat_2010_smc} and Large Magellanic
Clouds \citep[LMC,][]{lat_2010_lmc}, and M31, identified a simple power law
relation between star-formation and gamma-ray luminosity \citep{lat_2010_local_group}. On spatial scales within an individual galaxy,
resolved images of the LMC at GeV energies show that gamma-ray
emissivity per hydrogen atom, and
hence CR intensity, is greatest near the massive star-forming region
30 Doradus \citep{lat_2010_lmc}.


EGRET observations \citep{cillis_2005_egret_stack} yielded flux upper limits for a collection of
star-forming galaxies beyond the Local Group (typical limits of 3--5 $\times
10^{-8}$ ph cm$^{-2}$ s$^{-1}$ in the $>0.1$ GeV energy range), and stacking searches for a collective
signal from the same galaxies produced no significant
detection. \cite{lenain_2011_local_group} recently reported flux upper limits in the 0.2--200 GeV energy range
for several galaxies located within 5 Mpc including M81, M83, IC 342, Maffei 1,
Maffei 2, and M94 using 29 months of data collected by the Large Area
Telescope (LAT) on board the \textit{Fermi Gamma-ray Space Telescope}
(\textit{Fermi}). The MAGIC Collaboration has reported a flux upper
limit in the energy range $\gtrsim$160 GeV towards the nearest ULIRG,
Arp 220 \citep[15 hours,][]{albert_2007_arp_220}, and
H.E.S.S. observations of NGC 1068 produced a flux upper limit above
210 GeV \citep[4.3 hours,][]{aharonian_2005_hess_ngc_1068}. 


In this paper, we use three years of \textit{Fermi} LAT data to
perform a systematic search for high energy gamma-ray emission from 64
star-forming galaxies beyond the Local Group selected on the basis of
their present star-forming activity. The next section describes our
galaxy sample. Section \ref{sec_observations_and_analysis} outlines our analysis
of LAT data and we present results in Section \ref{sec_results}, including
updated spectral energy distributions for significantly detected galaxies and flux upper limits for
the remaining candidates. Five Local Group galaxies previously studied in LAT
data, the SMC, LMC, Milky Way, M31, and M33, are additionally included in a population study of low-redshift
star-forming galaxies. Our primary objective is to explore the global
properties of galaxies related to their CR-induced emissions. We find
that a simple power law relationship between gamma-ray
luminosity and SFR reported for Local Group galaxies \citep{lat_2010_local_group} also
describes the larger set of star-forming galaxies examined here. The implications of this scaling relation both for the
physics of CRs and the contribution of non-AGN-dominated star-forming
galaxies to the isotropic diffuse gamma-ray background are discussed in
Section \ref{sec_discussion}, and we predict which galaxies might be detected over the course of a 10-year
\textit{Fermi} mission. A standard $\Lambda$CDM cosmology with
$\Omega_{M}=0.3$, $\Omega_{\Lambda}=0.7$, and $H_0=75$ km s$^{-1}$ Mpc$^{-1}$ is used throughout.

%% file: sec_candidate_selection.tex
\section{Galaxy Sample}\label{sec_candidate_selection}


Massive star-formation is fueled by the availablity of dense molecular
gas in the interstellar medium (ISM). In order to select a sample of
galaxies with unambiguous ongoing star formation, we base our sample
of galaxies on the HCN survey of \cite{gao_solomon_2004_hcn_survey}. HCN $J=1-0$ line emission is
stimulated in the presence of dense molecular gas ($n_{{\rm H}_2}> 3
\times 10^4$ cm$^{-3}$) typically associated with giant
molecular clouds where the majority of star formation occurs \citep{solomon_1979_gas}. The total quantity of molecular gas in the
dense phase as measured by HCN line emission exhibits a low-scatter linear
correlation with both total IR luminosity \citep{gao_solomon_2004_dense_molecular_gas} and
RC luminosity \citep{liu_2010_radio_ir_gas} and is considered a reliable indicator
of SFR. Observations of Galactic molecular clouds suggest that these
relations hold at the scale of individual molecular
cloud cores as well \citep{wu_2005_gas}. 

The HCN survey is the most complete study to date in terms of total dense molecular gas content, including
nearly all of the nearby IR-bright galaxies with strong CO emission in the northern
sky ($\delta\geq-35^{\circ}$).\footnote{The HCN survey includes galaxies with 60 $\mu$m/100 $\mu$m emission
larger than 50 Jy/100 Jy and CO line brightness temperatures larger
than 100mK for spirals or 20mK for LIRGs/ULIRGs.} Additional galaxies
with globally measured HCN emission data
available in the literature are also included in the sample. 
Objects at Galactic latitudes $|b|<10^{\circ}$ are excluded due to the
strong diffuse emission from the Galactic plane.
Our final candidate list of 64 galaxies beyond the Local Group includes more than a dozen large
nearby spiral galaxies, 22 LIRGs, and 9 ULIRGs.
Global properies of the galaxies including RC luminosity, total IR luminosity,
and HCN luminosity are provided in Table \ref{table_candidates}. 


Many of the IR-bright starburst galaxies in our candidate list have been previously suggested as
interesting targets for gamma-ray telescopes on the
basis that a significant fraction of CR protons may interact in dense
molecular clouds before escaping the ISM
\citep{voelk_1989_calorimeter,akyuz_1991_m82,paglione_1996_ngc_253,torres_2004_glast,thompson_2007_calorimeter}.
Furthermore, galaxies well-known to host non-thermal leptonic
populations are naturally included in this sample since diffuse radio emission
associated with synchrotron radiation correlates with IR luminosity
\citep{condon_1991_radio_ir,yun_2001_radio}.

 
Some of the galaxies in our sample host radio-quiet AGN with low-level
jet activity. The galaxies associated with sources in the \textit{Swift} BAT
58-month survey catalog (15--195 keV) with AGN-type designations
\citep{baumgartner_2010_swift} are listed in the rightmost column of Table
\ref{table_candidates}. Hard X-ray emission is a strong and relatively
unbiased identifier of AGN activity \citep{burlon_2011_swift_agn}, while soft
X-ray emission from AGN is more often obscured by intervening
circumnuclear and ISM material. X-ray emission from AGN does not require presence of a relativistic jet.


Five Local Group galaxies previously examined in LAT data are included in
the multiwavelength comparisons appearing in Section
\ref{subsec_multiwavelength}. Multiwavelength data for those galaxies are summarized
in Table \ref{table_local_group}. 


The largest redshift of any galaxy in our sample is $z\sim0.06$.

%% file: sec_observations_and_analysis.tex
\section{LAT Observations \& Data Analysis}\label{sec_observations_and_analysis}


The \textit{Fermi}-LAT is a pair-production telescope with large effective area ($\sim$8000 cm$^2$
on axis for $E >$1 GeV) and field of view ($\sim$2.4 sr
at 1 GeV), sensitive to gamma rays in the energy
range from 20 MeV to $> 300$ GeV. 
Full details of the instrument and descriptions of the on-board and
ground data processing are provided in \cite{lat_2009_instrument}, and
information regarding on-orbit calibration procedures is given by \cite{lat_2009_on-orbit}. The LAT normally operates in a scanning `sky-survey' mode which
provides coverage of the full sky every two orbits ($\sim$3
hours). For operational reasons, the standard rocking angle (defined as the angle between the
zenith and center of the LAT field of view) for survey mode was
increased from $35^{\circ}$ to $50^{\circ}$ on 3 Sep 2009. 


This work uses data collected in sky-survey mode from 2008 Aug 4 to
2011 Aug 4 for the analysis of 64 celestial $15^{\circ}\times15^{\circ}$
regions of interest (RoI) centered on the positions of the galaxies in
our sample. We accept only low-background
`source' class photon candidate events \citep{lat_2009_instrument}
corresponding to the P7V6 instrument
response functions with reconstructed energies 0.1--100 GeV. In order
to reduce the effects of gamma rays produced by CR interactions in the
upper atmosphere \citep{lat_2009_earth}, we discard photons arriving from
zenith angles $> 100^{\circ}$, exclude time periods when part of the
RoI was beyond the zenith angle limit, and also exclude time periods when the
spacecraft rocking angle exceeded $52^{\circ}$.


The data were analyzed using the LAT Science Tools software package
(version 09-25-02).\footnote{Information regarding the LAT
Science Tools package, diffuse models, instrument response functions,
and public data access is available from the \textit{Fermi} Science
Support Center (\url{http://fermi.gsfc.nasa.gov/ssc/}).} The model for
each celestial RoI contains templates for the
diffuse Galactic foreground emission ({\tt gal\_2yearp7v6\_v0.fits}), a spectrum for the isotropic
diffuse emission (composed of both photons and residual
charged particle background, {\tt iso\_p7v6source.txt}), and all individual sources
reported in the LAT 2-year source catalog \citep[2FGL,][]{lat_2012_2fgl}
within $12^{\circ}$ of the target galaxies. For the individual LAT
sources, we assumed spectral models and parameters reported
in the 2FGL catalog. The candidate sources corresponding to the galaxies in
our sample were modeled as point sources with power-law spectra,
$dN/dE\propto E^{-\Gamma}$, at the optically-determined positions of the galaxies. Due to their
typical angular sizes of $\mathcal{O}(0.1^{\circ})$ or smaller, and
fluxes close to or below the LAT detection threshold, the
galaxies considered in this work are not expected to be resolved as
spatially extended beyond the energy-dependent LAT point-spread
function \citep{lande_2012_extended}.

The normalization and photon index of each gamma-ray source candidate were fitted using a maximum likelihood
procedure suitable for the analysis of binned photon data ({\tt gtlike}). The photons
within each RoI were binned spatially into $0.1^{\circ}$-sized pixels
and into 30 energy bins uniformly spaced in log-energy (10 bins for
each power of 10 in energy). During the maximum likelihood fitting,
the normalizations of the diffuse components were left free. We also fit the
normalizations of all neighboring LAT sources
within 4$^{\circ}$ of the target galaxy positions, or located within
the $15^{\circ}\times15^{\circ}$ RoI and detected with exceptionally
high significance in the 2FGL catalog\footnote{More than 1000
attributed photons in the 2FGL dataset or $TS>500$; see Section
\ref{subsec_lat_results} for the definition of $TS$ value}.

%% file: sec_results.tex
\section{Results}\label{sec_results}

Results from the analysis of 64 galaxies beyond the
Local Group in LAT data are presented first, followed by a discussion of
multiwavelength relationships using the full sample of 69 star-forming galaxies.

\subsection{LAT Data Analysis Results}\label{subsec_lat_results}


Table \ref{table_results} summarizes results from the maximum
likelihood analyses. Source detection significance is determined using the
Test-Statistic ($TS$) value, $TS\equiv-2(\ln(L_{0})-\ln(L_{1}))$,
which compares the likelihoods of models
including the galaxy under consideration ($L_{1}$) and the
null-hypothesis of no gamma-ray emission from the galaxy
\citep[$L_{0}$,][]{mattox_1996_likelihood}. Significant ($TS$>25) gamma-ray excesses
above background were detected in directions coinciding with four
galaxies in the sample: two prototypical starburst galaxies, M82 and NGC 253,
and two starburst galaxies also containing Seyfert 2 nuclei, NGC 1068, and NGC
4945. Each of the four gamma-ray sources is associated with the
corresponding galaxy from our sample according to the 2FGL catalog \citep{lat_2012_2fgl}.

We additionally obtain relatively large $TS$ values in the
directions of NGC 2146 ($\sim20$) and M83 ($\sim15$). These excesses do
not pass the conventional threshold of $TS>25$ required to claim a
source detection. \cite{lenain_2011_local_group} noted an excess in the
vicinity of M83, but determined that the best-fit position was
inconsistent with the galaxy location and instead proposed an association with the
blazar 2E 3100. We will return to these two galaxies in Section \ref{subsec_predictions}. 


CR-induced gamma-ray emission from galaxies is expected to be steady on the
timescales of our observations. Variability is tested for the four significantly detected
sources by partitioning the full observation period into 12 time
intervals of $\sim90$ days each and performing a separate
maximum-likelihood fit for each of the shorter time-periods. The
resulting lightcurves are shown in Figure \ref{fig_lightcurve}. None of
the four sources show significant changes in flux over the three years of LAT
observations. This finding is consistent with the results
obtained by \cite{lenain_2010_seyfert} for NGC 1068 and NGC 4945.  


Spectral energy distributions of M82, NGC 253, NGC 1068, and NGC 4945
are shown in Figure \ref{fig_sed}. Each flux measurement represents a
separate maximum-likelihood fit following the procedures described
above, but for 6 logarithmically spaced energy bins. Each energy bin
is further divided into 5 logarithmically-spaced sub-bins of energy for binned
likelihood analysis to maintain $\sim10$ bins for each power of
10 in energy. Using the power-law spectral models, the maximum likelihood photon index
values for the four galaxies are in the range 2.1--2.4. For M82 and
NGC 253, the extrapolated power-law fits from the
LAT energy range naturally connect with spectral measurements obtained by
imaging air-Cherenkov telescopes at higher energies \citep{acciari_2009_m82,acero_2009_ngc_253,ohm_2011_ngc_253}. 




Integral flux upper limits in the range 0.1--100 GeV at the 95\%
confidence level are provided in Table \ref{table_results} for galaxies not significantly detected by the
LAT. Limits are computed using the profile likelihood technique, for
which the flux of a target source is varied over a range while
simultaneously maximizing the model likelihood with respect to all
other parameters. One-sided 95\% CL upper limits correspond to a
decrease in profile likelihood, $L_p$, of $2 \Delta \ln(L_p) = 2.71$
relative to the maximum likelihood. The flux upper limits
presented in Table \ref{table_results} are derived using a power-law spectral model with photon
index $\Gamma=2.2$, corresponding to the typical index of the four
LAT-detected starbursts. The upper limits increase by 10--20\% (less
constraining) on average when assuming a photon index of 2.3.

\subsection{Gamma Rays from Starbursting Seyfert 2 Galaxies}\label{subsec_seyfert}


NGC 1068 and NGC 4945 deserve special attention as galaxies with both
circumnuclear starbursts and radio-quiet obscured AGN. The gamma-ray spectra of
NGC 1068 and NGC 4945 are similar to those of M82 and NGC
253, and none of the four galaxies show indications of gamma-ray
variability. \cite{lenain_2010_seyfert} suggested a model for NGC 1068
in which gamma rays are produced mainly through inverse Compton scattering of IR photons by
electrons in the misaligned jet at distances $\sim0.1$ pc from the
central engine. In this scenario, NGC 1068 is viewed as the first member
of a new class of gamma-ray sources.

 
Studies of X-ray selected Seyfert galaxies using LAT data demonstrate
that Seyfert galaxies hosting radio-quiet AGN are generally gamma-ray
quiet as a population \citep{teng_2011_seyfert,lat_2012_seyfert}. Excepting NGC 1068 and NGC 4945, no other radio-quiet
Seyfert galaxies have been detected by the LAT. The upper limits in
0.1--100 GeV luminosity inferred for several other nearby Seyfert galaxies are below the gamma-ray
luminosity of NGC 1068 \citep{lat_2012_seyfert}. The tightest
constraint is found for NGC 4151, located
at a distance of 11.2 Mpc, with gamma-ray luminosity $<2 \times
10^{40}$ erg s$^{-1}$ (a factor $\sim10$ below that of NGC 1068).



The relative contributions of CR interactions versus AGN activity to the gamma-ray emission of NGC 1068 and
NGC 4945 is not yet definitively established. In the multiwavelength
analysis which follows, we will consider both the full sample of
analyzed galaxies, and a subsample with galaxies hosting \text{Swift}-BAT-detected AGN removed.

\input{subsec_multiwavelength.tex}

%% file: subsec_multiwavelength.tex
\subsection{Multiwavelength Luminosity Comparisons}\label{subsec_multiwavelength}


Many authors have proposed scaling relationships between galaxy
star-formation rates (SFRs) and gamma-ray
luminosities motivated by the connections between interstellar
gas, star formation, supernovae, and CRs \citep[e.g.][]{pavlidou_2002_guaranteed,torres_2004_glast,thompson_2007_calorimeter,stecker_2007_starburst,persic_2010_cr_energy_density,lacki_2011_starbursts_interpretation}. We now
examine relationships between gamma-ray luminosity and several
photometric tracers of star formation taking advantage of the
three-year accumulation of LAT data. 


The radiative output of star-forming galaxies across much of the
electromagnetic spectrum is related to the abundance of short-lived
massive stars. Many photometric estimators of the recent star
formation histories of galaxies have been used. Total IR luminosity 8--1000 $\mu$m is one well-established tracer
of the SFR for late-type galaxies \citep[reviewed
by][]{kennicutt_1998_hubble}. The conversion proposed by \cite{kennicutt_1998_schmidt}, 

\begin{equation}
\frac{\rm SFR}{{\rm M}_{\odot}\:{\rm yr}^{-1}}=\epsilon 1.7 \times 10^{-10} \frac{L_{8-1000
\; \mu{\rm m}}}{L_{\odot}},
\label{eq_sfr_ir}
\end{equation}

\noindent assumes that thermal emission of interstellar dust
approximates a calorimetric measure of radiation produced by young (10--100 Myr) stellar
populations. All SFRs quoted in this work consider the stellar mass
range 0.1--100 $M_{\odot}$. The factor $\epsilon$ depends on the assumed initial mass
function (IMF), with $\epsilon=1$ for the \cite{salpeter_1955_imf} IMF
originally used by \cite{kennicutt_1998_schmidt}. We will use
$\epsilon=0.79$ to convert\footnote{We use the conversion factor proposed by
\cite{crain_2010_imf_conversion} for SFRs estimated from total IR
luminosity: ${\rm SFR}_{\rm Chabrier}=0.79 \; {\rm SFR}_{\rm
Salpeter}$.} to the \cite{chabrier_2003_imf} IMF, proposed after further
studies of star formation in the 0.1--1 $M_\odot$ mass range.

Several other photometric SFR estimators have been calibrated using
the \cite{kennicutt_1998_schmidt} total IR luminosity relation. These
include RC luminosity at 1.4 GHz produced by synchrotron-emitting
CR electrons \citep{yun_2001_radio},

\begin{equation}
{\rm SFR}({\rm M}_{\odot}\:{\rm yr}^{-1})= \epsilon (5.9\pm1.8) \times
10^{-22} L_{1.4 \; {\rm GHz}},
({\rm W\:Hz}^{-1}),
\label{eq_sfr_rc}
\end{equation}

\noindent and HCN $J$=1-0 line luminosity indicating the quantity of dense molecular
gas available to form new stars \citep{gao_solomon_2004_dense_molecular_gas},

\begin{equation}
{\rm SFR}({\rm M}_{\odot}\:{\rm yr}^{-1})= \epsilon 1.8 \times 10^{-7} L_{{\rm HCN}} (\rm{K\:km\:}\rm{s}^{-1}\:{\rm pc}^{2}).
\label{eq_sfr_hcn}
\end{equation}

\noindent Although these three SFR estimators are intrinsically linked, each explores a different
stage of stellar evolution and is subject to different astrophysical and
observational systematic uncertainties.


Figures \ref{fig_rc_correlation} and \ref{fig_ir_correlation} compare the gamma-ray luminosities of 
galaxies in our sample to their differential luminosities at 1.4 GHz, and total IR
luminosities (8--1000 $\mu$m), respectively. A second abscissa axis
has been drawn on each figure to indicate the
estimated SFR corresponding to either RC or total IR luminosity using
equations \ref{eq_sfr_rc} and \ref{eq_sfr_ir}. The upper panels of
Figures \ref{fig_rc_correlation} and \ref{fig_ir_correlation} directly
compare luminosities between wavebands, whereas the lower panels
compare luminosity ratios. Taken at face value,
the two figures show a clear positive correlation between gamma-ray
luminosity and SFR, as has been reported previously in LAT data
\citep[see in this context][]{lat_2010_local_group}. However,
sample selection effects, and galaxies not yet detected in gamma rays
must be taken into account to properly determine
the significance of the apparent correlations.





We test the significances of multiwavelength correlations using the modified Kendall $\tau$
rank correlation test proposed by
\cite{akritas_siebert_1996_kendall}. This method is an example of
`survival analysis' techniques, suitable for the analysis of
partially-censored datasets (i.e. containing a mixture of detections and
upper limits). The Kendall $\tau$ coefficient, $\tau\in[-1,1]$, indicates the degree of
positive or negative correlation between two quanitites by comparing
the ordering of each pair of points in the dataset. For example, $\tau=1$ describes
a set of uncensored data points which are monotonically
increasing. This non-parametric approach makes no assumption regarding
the particular mathematical form for the relationship between compared quantities. 
See Appendix \ref{app_sec_kendall_tau} 
for a detailed description of the method.


The multiwavelength relationships are tested in luminosity-space
because we are primarily interested in the intrinsic galaxy
properties, and importantly, because multiwavelength comparisons in
flux-space can either falsely suggest a non-existant correlation or
obscure a genuine physical correlation if the underlying relationship
between intrinsic luminosities is non-linear \citep[][ and see
Appendix
\ref{app_sec_flux_hazard}]{feigelson_berg_1983_3cr,kembhavi_feigelson_1986_core}.
We performed a series of Monte Carlo simulations to investigate potential biases resulting from
selection effects for flux-limited samples of objects compared in
luminosity-space. The generalized Kendall $\tau$
correlation test is found to be robust against the selection effects expected for
our application. A brief discussion of results from that study is
presented in Appendix \ref{app_sec_simulations}.


Significances of the multiwavelength relationships are computed by
comparing the $\tau$ correlation coefficient of the actual data to the
distribution of $\tau$ correlation coefficients which could be
obtained under the null hypothesis of independence between
wavebands. Null hypothesis datasets are generated by scrambling derived gamma-ray
luminosities among galaxies. Only `observable'
permutations are retained to account for the truncation of measurable
gamma-ray luminosities imposed by the LAT flux sensitivity (following
the method of \cite{efron_petrosian_1999_truncated}). Specifically, we require that
the resultant gamma-ray flux (or upper limit) of each galaxy in a scrambled
data set could have been measurable by the LAT in order for the scrambled data set to be included in the
null hypothesis distribution. In practice, each permutation
represents an exchange of gamma-ray luminosities between two randomly
selected galaxies so that statistically meaningful null hypothesis
distributions can be efficiently generated. Gamma-ray flux sensitivity
thresholds are determined by randomly sampling from the distribution of gamma-ray flux
upper limits presented in Table \ref{table_results} for each galaxy in
the pair to be exchanged. We perform 1000 such exchanges on the actual data before counting
permutations towards the null hypothesis distribution. 


Inaccuracies in the distance measurements to the galaxies are included
in our analysis because these errors are propagated in the transformation of measured fluxes into
luminosities. Scatter in distance measurements applied to a collection of objects tends to
broaden the distribution of objects in luminosity-space, and can therefore induce
positive correlation. For the galaxies in our sample at distances 1 Mpc$\lesssim D \lesssim$100 Mpc,
the typical distance measurement precision for individual galaxies is
10--20\% \citep[][ and references therein]{freedman_2001_hubble}. For
our particular sample, we can compare the distances provided
in the HCN survey of \cite{gao_solomon_2004_hcn_survey} to
those reported in the IRAS Revised Bright Galaxies Sample
\citep{sanders_2003_iras_rbgs}, and find that for galaxies beyond the
Local Group, the average discrepancy between
reported distances is 10\%. We assume that measurement errors are
normally distributed and centered on the true distances to the galaxies.


Figure \ref{fig_kendall} shows an example of the correlation significance
test applied to the relationship between gamma-ray
and IR luminosity using the full sample of star-forming
galaxies. The distribution of correlation coefficients for $10^6$
`observable' permutations of the actual data are plotted for 3
levels of distance measurement scatter. The null hypothesis
distributions tend towards positive values due to the truncation of
measureable gamma-ray luminosities, effectively requiring a larger correlation
coefficient for the actual data in order to claim a
significant relationship between wavebands. As the level of scatter in
distance measurements is increased, the tails of null hypothesis
distributions extend to larger positive values. The probability
that a null hypothesis data set would have a correlation coefficient larger than that
of the actual data corresponds to the $P$-value measure of correlation
significance.


Given uncertainties for the measured gamma-ray fluxes of galaxies in our sample,
the correlation coefficient of the actual data is more
appropriately viewed as a probability density as opposed to a single
value (the Kendall $\tau$ statistic does not explicitly include
measurement errors). To account for this uncertainty, we draw a flux value for
each LAT-detected galaxy from a normal distribution centered on the
best-fit flux with standard deviation equal to the reported flux
uncertainty. A correlation coefficient is then computed using the sampled gamma-ray fluxes. This process is repeated many times to generate a
distribution of correlation coefficients representing the actual data,
as shown by the gray bands in Figure \ref{fig_kendall}. Finally, we integrate over
the probability density of correlation coefficients representing the actual
data to obtain significances for the multiwavelength relations.


The results of the correlation tests are summarized in Table
\ref{table_significance}. Using the full sample of 69 star-forming
galaxies, the null hypothesis of independence between gamma-ray
luminosity and either RC or total IR luminosity can be rejected
at the $\gtrsim$99\% confidence level allowing for 20\% uncertainty on
the distance measurements to the galaxies. No evidence for a correlation between
gamma-ray luminosity and HCN line luminosity is found in this study, an apparent discrepancy considering the
previously established correlations between RC, IR, and HCN line
luminosities \citep{liu_2010_radio_ir_gas}. However, global HCN line
luminosity estimates are not available for the SMC and LMC, M31, and
M33, which effectively reduces the luminosity range over which the correlation
between wavebands can be tested, and excludes three LAT-detected
galaxies from the analysis.


We also considered the more conservative case in which galaxies hosting
\textit{Swift}-BAT detected AGN, including NGC 1068 and NGC 4945 along
with seven others, were removed from the sample given the potential
contributions of the AGN to the broad-band emission of those
galaxies. In that case, the $P$-values for the
correlations between gamma-ray luminosity and
either RC or IR luminosity are $\sim0.05$ allowing for
20\% uncertainty on the distance measurements. 


Even in the most conservative case described above, the data prefer a correlation between gamma-ray luminosity and both RC and IR
luminosity. Although strong conclusions regarding the significance of the
multiwavelength correlations cannot be made at the time, at minimum, scaling
relations derived from the current sample have predictive value and
offer a testable hypothesis for further observations. 


We fit scaling relationships between wavebands
using simple power law forms. For example,

\begin{equation}
\log \left(\frac{L_{0.1-100 \; {\rm GeV}}}{{\rm erg \; s}^{-1}}\right) =
\alpha \log \left(\frac{L_{8-1000 \; \mu{\rm m}}}{10^{10} L_\odot}\right)
+ \beta,
\label{eq_scaling_ir}
\end{equation}

\noindent parameterizes the relationship between gamma-ray and total IR luminosity.

Two regression methods are employed: the Expectation-Maximization (EM)
algorithm, and the Buckley-James algorithm. The EM algorithm is
similar to the least-squares fitting
method, and assumes that the intrinsic residuals of gamma-ray
luminosity are normally distributed in log-space about the
regression line for fixed values of either RC or IR
luminosity. Non-detections are incorporated in the regression analysis
by determining the degree to which the upper limits are compatible
with the assumed dispersion about the regression line. The
variance of the intrinsic residuals becomes a third parameter in the fit.
The Buckley-James algorithm is similar in approach, but generalizes to
cases for which the intrinsic distribution of residuals relative to the regression
line is not normally distributed. Instead, the
intrinsic dispersion is estimated using the Kaplan-Meier distribution
obtained from the scatter within the dataset itself. \cite{isobe_1986_censored} provide
details regarding the theory and implementation several commonly used survival analysis
methods including the EM and Buckley-James regression algorithms.

Table \ref{table_regression} reports the best-fits of gamma-ray
luminosity versus either RC luminosity or total IR luminosity, which
were evaluated using the {\tt ASURV} Rev 1.2
code \citep{lavalley_1992_asurv}.\footnote{{\tt ASURV} and
other tools for the analysis of censored data sets are available from
the Penn State Center for Astrostatistics
(\url{http://www.astrostatistics.psu.edu/statcodes/sc\_censor.html}).}
The two regression methods yield consistent results. For both the RC
and total IR tracers of SFR, a nearly linear power law index of
$\alpha$=1.0-1.2 is preferred.
The scaling indices between gamma-ray luminosity and SFR tracers found
for the larger sample of galaxies is consistent within
uncertainties with that obtained considering the Local Group galaxies
alone, $\alpha$=1.4$\pm$0.3 \citep{lat_2010_local_group}. In
particular, the gamma-ray luminosity upper limits found for LIRGs
and ULIRGs disfavor a strongly non-linear scaling relation. 
Table \ref{table_regression} also contains regression
parameters fitted after removing galaxies hosting \textit{Swift}-BAT
detected AGN from the sample, which would be appropriate if
the broadband emissions of those galaxies were heavily influenced by
AGN activity. The fitted parameters in both cases are consistent within
statistical uncertainties.

Note that the intrinsic dispersion values presented in Table
\ref{table_regression} should be viewed as upper
limits because we have not made an attempt to account for measurement uncertainties
in the gamma-ray fluxes.

The best-fit power laws obtained using the EM algorithm are plotted in
Figures \ref{fig_rc_correlation} and \ref{fig_ir_correlation}. The darker
shaded regions represent uncertainty in the fitted parameters from the
regression and the lighter shaded regions indicate the addition of intrinsic residuals to one standard
deviation. None of the limits from non-detected galaxies are in
strong conflict with the best-fit relations. 

Luminosity ratios of the form shown in the lower panels of Figures
\ref{fig_rc_correlation} and \ref{fig_ir_correlation} have a direct physical meaning in terms of
the energy radiated in different parts of the electromagnetic
spectrum. For a galaxy with non-thermal
emission powered mainly by CR interactions, and having a SFR of 1
${\rm M}_{\odot}$ yr$^{-1}$ (similar to the Milky Way), the
corresponding luminosity ratios between wavebands are 

\begin{equation}
\log \left(\frac{L_{0.1-100 \; \rm{GeV}}}{L_{1.4 \; \rm{GHz}}}\right) = 1.7
\pm 0.1_{\rm (statistical)} \pm 0.2_{\rm (dispersion)},
\end{equation}

\noindent and

\begin{equation}
\log \left(\frac{L_{0.1-100 \; \rm{GeV}}}{L_{8-1000 \; \mu\rm{m}}}\right) = -4.3 \pm 0.1_{\rm
(statistical)} \pm 0.2_{\rm (dispersion)},
\end{equation}

\noindent found using the full sample of galaxies with the EM regression algorithm.

Further gamma-ray observations are required to conclusively
establish the multiwavelength correlations examined in this
section. Additional data may also show that scaling relations beyond
simple power law forms are required once farther and more active
sources are either detected or are constrained by more stringent
gamma-ray upper limits.

%% file: sec_discussion.tex
\section{Discussion}\label{sec_discussion}


In recent years, increasingly sensitive observations of external galaxies at both GeV
and TeV energies have shown
that starburst galaxies such as M82 and NGC 253
are characterized by harder gamma-ray spectra relative to quiescent
galaxies of the Local Group
\citep{acero_2009_ngc_253,acciari_2009_m82,lat_2010_starburst,ohm_2011_ngc_253}, and
that global gamma-ray luminosities of galaxies likely scale quasi-linearly with
SFRs \citep[e.g.][]{lat_2010_starburst,lat_2010_local_group,lenain_2011_local_group}.

Figure \ref{fig_sed_all} compares the spectra
of eight star-forming galaxies detected by gamma-ray telescopes. The more
luminous galaxies also have comparatively harder gamma-ray
spectra. Whereas the power-law spectral indices of M82 and NGC 253 are 2.2--2.3
extending to TeV energies, the spectra of the LMC and SMC steepen above $\sim2$ GeV
\citep{lat_2010_smc,lat_2010_lmc}. 

Scaling relations of the type examined in Section
\ref{subsec_multiwavelength}, and the observed gamma-ray spectra of
star-forming galaxies, have implications both for the physics of CRs
in the ISM and for the contribution of star-forming galaxies to
the isotropic diffuse gamma-ray emission. In this section, we
concentrate on the global non-thermal radiation features of star-forming
galaxies from a population standpoint.

\input{subsec_cr.tex}

\input{subsec_egb.tex}

\input{subsec_predictions.tex}

%% file: subsec_cr.tex
\subsection{Physics of Cosmic Rays in Star-forming Galaxies}\label{subsec_cr}


A discussion of scaling relations for diffuse gamma-ray radiation
from star-forming galaxies invites comparison with the extensively
studied RC-IR correlation. The RC-IR correlation extends over multiple
galaxy types including irregulars, spirals, and ellipticals with
on-going star formation
\citep{de_jong_1985_radio_ir,wunderlich_1987_radio_ir,dressel_1988_radio_ir,wrobel_1988_radio_ir,wrobel_1991_radio_ir},
covers galaxies with magnetic energy densities likely ranging at least 4 orders of
magnitude \citep{condon_1991_radio_ir,thompson_2006_magnetic_field}, and applies to
galaxies out to at least $z\sim1$ \citep{appleton_2004_radio_ir}. High-resolution imaging has also shown the
correlation to hold on $\sim$100 pc scales within individual galaxies
\citep{marsh_1995_radio_ir,paladino_2006_radio_ir,murphy_2006_radio_ir}.


A simple model proposed to explain the RC-IR correlation posits that
galaxies are effective `calorimeters' of both UV photons and CR
electrons \citep{voelk_1989_calorimeter}. In this picture, if the ISM is
optically thick to UV photons, reprocessed starlight emitted
in the IR becomes a proxy for the abundance of
massive stars, and hence SFR. The luminosity of CR electrons
is assumed to be proportional to the SFR in the basic calorimeter
model. If either synchrotron losses dominate for CR electrons, or if
the ratio of synchrotron losses to other loss-mechanisms is
uniform for many galaxy types, then the RC luminosity should also increase
together with the SFR. The latter possibility requires some degree of
fine-tuning \citep[but see][]{lacki_2010_conspiracy}. Numerical models of CR propagation in the Milky Way
suggest that our Galaxy is an effective CR electron calorimeter \citep{strong_2010_global}.


Gamma-ray observations extend our study of CRs to include hadronic
populations beyond the solar system, and notably, to measure the
large-scale propagated spectra of CR nuclei in the Milky Way and other
galaxies. At energies above the threshold for pion production,
hadronic gamma rays have the same spectral index
as the underlying CR proton spectrum in the thin-target regime
relevant for proton-proton interactions in the ISM. Diffuse pionic
gamma rays presumably offer direct access to the majority of CR energy
content, although leptonic diffuse
emission and individual sources within galaxies must also contribute
to the total emission. The gamma-ray spectra of the SMC and LMC are both
consistent with models in which pionic gamma-rays dominate the high
energy emission \citep{lat_2010_smc,lat_2010_lmc}. Many authors have
argued that the majority of starburst emission at energies $>0.1$
GeV is produced by interactions of CR nuclei in the interstellar
medium, assuming primary CR proton to electron ratios similar
to values found in the Milky Way \citep[e.g.][]{paglione_1996_ngc_253,torres_2004_arp_220,persic_2008_m82,lacki_2010_conspiracy}.


For the Milky Way in particular, global CR-induced gamma-ray emission above 0.1 GeV stems mostly
from neutral pion decay \citep{bloemen_1985_diffuse,bertsch_1993_diffuse}, with
leptonic processes contributing 30--40\% in the energy range 0.1--100
GeV \citep{strong_2010_global}. The diffuse neutral-pion-decay
component for the Milky Way has spectral index $\sim2.75$ above $\sim1$ GeV,
matching the locally measured CR proton spectrum \citep{simpson_1983_cr_composition,sanuki_2000_bess,adriani_2011_cr_proton_helium}. CR escape from the Milky Way plays
a major role in the formation of
the measured CR spectrum, but other galaxies with higher
concentrations of ambient gas may work differently,
behaving as calorimeters of CR nuclei, as suggested by
\cite{thompson_2007_calorimeter}. Approximately 10\% (possibly up to
20\%) of the global gamma-ray emission of the
Milky Way is thought to originate from individual sources \citep{strong_2007_population_synthesis}.


We now review models for the interactions of CR nuclei to examine
physical quantities involved in the scaling of hadronic gamma-ray
emission, following the formalism of recent works by
\cite{persic_2010_cr_energy_density} and
\cite{lacki_2011_starbursts_interpretation}. Neutral pion decay emission is related both to the
energy density of CR nuclei, $U_{\rm p}$, and to the ambient gas density,
$n$, which serves as target material for inelastic collisions. The
total hadronic gamma-ray luminosity (erg s$^{-1}$) can be expressed as 

\begin{equation}
L_{\gamma,\pi^{0}} = \int_V\int E_{\gamma}\frac{dq}{dE_{\gamma}}\;n(\vec{r})\;U_{\rm p}(\vec{r})dE_{\gamma}dV,
\label{eq_luminosity_general}
\end{equation}

\noindent integrating over gamma-ray energies, $E_{\gamma}$, and over the whole galactic volume,
$V$, which comprises the disk and surrounding halo occupied by CRs. The factor
$q(\ge E_{\gamma})$ denotes gamma-ray emissivity (s$^{-1}$ H-atom$^{-1}$
eV$^{-1}$ cm$^3$) normalized to the CR energy density
\citep{drury_1994_cosmic_rays}. Next, we define an average
\textit{effective} density of ambient gas encountered by CR nuclei:

\begin{equation}
\langle n_{\rm eff}\rangle = \frac{\int_V n(\vec{r})U_{\rm p}(\vec{r})
dV}{\int_V U_{\rm p}(\vec{r}) dV}.
\end{equation}

\noindent Note that $\langle n_{\rm eff}\rangle$ can be substantially
lower than the average gas density found in the galaxy disk. CR nuclei
 of the Milky Way are thought to spend most of their time in the
 low-gas-density halo, based upon combined measurements of unstable
 isotopes and spallation products (e.g. the B/C ratio) in the locally-observed CR composition
 \citep[reviewed by][]{strong_2007_cr_review}. Using the above definition
 for the average effective gas density, equation \ref{eq_luminosity_general} separates to

\begin{equation}
L_{\gamma,\pi^{0}} = \int_V U_{\rm p} dV \times \langle n_{\rm eff}\rangle \times \int E_{\gamma}\frac{dq}{dE_{\gamma}} dE_{\gamma}.
\label{eq_luminosity_effective}
\end{equation}

The leading factor of equation \ref{eq_luminosity_effective}
represents the total energy of CR nuclei, and can be re-written as the product of the total luminosity of CR
nuclei, $L_{\rm p}$, and their characteristic residence time in the
galactic volume, $\tau_{\rm res}$:

\begin{equation}
\int_V U_{\rm p} dV = L_{\rm p}\;\tau_{\rm res}.
\label{eq_cr_energy_density}
\end{equation}

The trailing factor of equation \ref{eq_luminosity_effective} depends
upon the parent spectrum of CR nuclei, for which we assume a power law spectral
distribution, $dN_{\rm p}/dK_{\rm p} \propto (K_{\rm p} / 1 \; {\rm
GeV})^{-\Gamma_{\rm p}}$, where $K_{\rm p}$ is the kinetic energy. Gamma-ray emissivities (s$^{-1}$
H-atom$^{-1}$ cm$^3$) in the energy range 0.1--100 GeV, 

\begin{equation}
\mathcal{Q}_{0.1-100 \; {\rm GeV}} = \int_{0.1 \; {\rm GeV}}^{100 \; {\rm GeV}}
E_{\gamma}\frac{dq}{dE_{\gamma}} dE_{\gamma},
\label{eq_emissivity}
\end{equation}


\noindent are computed
using the proton-proton interaction cross section parametrizations of
\cite{kamae_2006_hadronic}. Note that the emissivities are normalized to the average energy
density of CR nuclei ($1 < K_{\rm p} < 10^6$ GeV) within the entire
galactic volume. A nuclear enhancement factor of 1.85 is included to account for heavy
nuclei in both CRs and target matter
\citep{mori_2009_enhancement}. The emissivities for $\Gamma_{\rm p} = 2-3$ range from $\mathcal{Q}_{0.1-100 \; {\rm GeV}} = (5.8-7.9) \times
10^{-17}$ s$^{-1}$ H-atom$^{-1}$ cm$^3$. 

Substituting equations \ref{eq_cr_energy_density} and
\ref{eq_emissivity} into equation \ref{eq_luminosity_effective} yields

\begin{equation}
L_{0.1-100 \; {\rm GeV},\pi^{0}} = L_{\rm p}\;\tau_{\rm res}\;\langle n_{\rm eff}\rangle\;\mathcal{Q}_{0.1-100 \; {\rm GeV}}.
\end{equation}

In the paradigm that SNRs are the primary sources of galactic CRs \citep{ginzburg_1964_origin_of_crs}, CR luminosity is
the product of the supernova rate, $\Gamma_{\rm SN}$,  the
kinetic energy released per supernova, $E_{\rm SN}$, and the fraction
of kinetic energy going into CR nuclei with kinetic energies $>1$ GeV
(i.e. acceleration efficiency), $\eta$:

\begin{equation}
L_{\rm p} = \Gamma_{\rm SN}\;E_{\rm SN}\;\eta. 
\end{equation}

\noindent Core-collapse supernova rates can be estimated
from SFRs given an initial mass function and a minimum stellar mass
required to produce a core-collapse supernova, which we
take to be 8 ${\rm M}_{\odot}$. For a \cite{chabrier_2003_imf} initial mass function, the
transformation between SFR in the mass range 0.1--100 ${\rm M}_{\odot}$ and core-collapse
supernova rate is $\Gamma_{\rm SN} \Psi = 
{\rm SFR}$, with $\Psi = 83 {\rm M}_{\odot}$. The average kinetic energy released per core-collapse supernova is
$E_{\rm SN}\sim10^{51}$ erg \citep{woosley_1995_sne}. The gamma-ray
luminosity becomes

\begin{equation}
L_{0.1-100 \; {\rm GeV},\pi^{0}} = \Psi^{-1}\;{\rm SFR}\;E_{\rm SN}\;\eta\;\tau_{\rm res}\;\langle n_{\rm eff}\rangle\;\mathcal{Q}_{0.1-100 \; {\rm GeV}}.
\end{equation}

\noindent Greater physical intuition can be realized by inserting
values for the physical quantities similar to those of the Milky Way,


\begin{align}
L_{0.1-100 \; {\rm GeV},\pi^{0}} = 9.4 \times 10^{38} \rm{erg\;s}^{-1}
\left(\frac{\rm SFR}{{\rm M}_{\odot}{\rm yr}^{-1}}\right) \notag\\
\left(\frac{\mathit{E}_{\rm SN}}{10^{51} \rm{erg}}\right) 
\left(\frac{\eta}{0.1}\right)
\left(\frac{\tau_{\rm res}}{10^7 {\rm yr}}\right)
\left(\frac{\langle n_{\rm eff}\rangle}{{\rm cm}^{-3}}\right),
\label{eq_luminosity_scaling}
\end{align}

\noindent assuming a power law index for CR nuclei of $\Gamma_{\rm p}
= 2.75$ corresponding to $\mathcal{Q}_{0.1-100 \; {\rm GeV}} = 7.8 \times
10^{-17}$ s$^{-1}$ H-atom$^{-1}$ cm$^3$.

A consistency check for this simple approach can be performed for the specific case of the Milky Way,
for which the column density of material traversed by CR nuclei, or `grammage,' is well
known \citep[e.g.][]{dogiel_2002_cr_luminosity}. The mean escape
length for CR nuclei in the Milky Way is $\bar{x} = \tau_{\rm res}\;m_{\rm p}\;\langle n_{\rm eff}\rangle\;c \approx 12$
g cm$^{-2}$ \citep{jones_2001_grammage} (cf. the interaction length is
$\sim$55 g cm$^{-2}$). We may equivalently express equation
\ref{eq_luminosity_scaling} in terms of the escape length as


\begin{align}
L_{0.1-100 \; {\rm GeV},\pi^{0}} = 6.0 \times 10^{38} \rm{erg\;s}^{-1}
\left(\frac{\rm SFR}{M_{\odot}{\rm yr}^{-1}}\right) \notag\\
\left(\frac{\mathit{E}_{\rm SN}}{10^{51} \rm{erg}}\right)
\left(\frac{\eta}{0.1}\right)
\left(\frac{\bar{x}}{10\;{\rm g\;cm}^{-2}}\right).
\label{eq_luminosity_scaling_grammage}
\end{align}

\noindent If we take SFR = 1.6 ${\rm M}_{\odot}{\rm yr}^{-1}$ corresponding
to a SN rate of $1.9 \; (\pm 1.1)$ per century \citep{diehl_2006_sn_rate}, then
our estimate for the pionic gamma-ray luminosity of the Milky Way becomes

\begin{equation}
L_{0.1-100 \; {\rm GeV},\pi^{0}} = 1.1 \times 10^{39} \rm{erg\;s}^{-1}
\left(\frac{\mathit{E}_{\rm SN}}{10^{51} \rm{erg}}\right)
\left(\frac{\eta}{0.1}\right).
\end{equation}

\noindent For comparison, the luminosity estimated from a full
numerical treatment (\texttt{GALPROP} code) incorporating an
ensemble of multiwavelength and CR measurements is $L_{0.1-100 \; {\rm
GeV},\pi^{0}} = (5-7) \times 10^{38}$ erg s$^{-1}$, with $E_{\rm SN} \eta =
(0.3-1) \times 10^{50}$ erg \citep{strong_2010_global}.

It is commonly assumed that the kinetic energy released per supernova, and
acceleration efficiency are universal from galaxy to
galaxy. In that case, the model expressed in equations
\ref{eq_luminosity_scaling} and \ref{eq_luminosity_scaling_grammage} implies that the hadronic gamma-ray
luminosity of the interstellar medium scales linearly with the SFR,
and with the product of residence time and effective ambient gas density (or
equivalently, grammage). These quantities are likely correlated in real galaxies. For example, several authors have argued that SFRs and gas
densities are related (e.g. via the Schmidt-Kennicutt law) so that a
non-linear scaling of gamma-ray luminosities with SFRs should be
expected: $L_{0.1-100 \; {\rm GeV},\pi^{0}} \propto {\rm SFR}^{1.4-1.7}$ \citep{persic_2010_cr_energy_density,fields_2010_galaxies_egb}.

In general, the residence time of CR nuclei in a galaxy, $\tau_{\rm res}$, can be
expressed as 

\begin{equation}
\tau_{\rm res}^{-1}=\tau_{\rm esc}^{-1}+\tau_{\rm pp}^{-1}. 
\end{equation}

\noindent A galaxy becomes a `calorimeter' of CR nuclei when the residence time
approximately equals the collisional energy loss timescale, $\tau_{\rm
res}\approx  \tau_{\rm pp}$, i.e. energy losses are dominated by inelastic collisions with
interstellar matter rather than by escape of energetic
particles. The proton-proton collisional energy loss timescale depends primarily on the
average gas density encountered by CR nuclei, and is nearly
independent of energy because the inelastic cross-section, $\sigma_{\rm pp}$, is only weakly
energy-dependent for CRs with kinetic energies greater than $\sim1$ GeV. The
collisional energy loss timescale is \citep{mannheim_1994_cr_interactions}

\begin{equation}
\tau_{\rm pp} \approx \left( 0.65\langle n_{\rm eff}\rangle\;\sigma_{\rm
pp}\;c \right)^{-1} \approx 5 \times 10^7 {\rm yr}\;\left(\frac{\langle n_{\rm eff}\rangle}{{\rm cm}^{-3}}\right)^{-1}. 
\end{equation}

By contrast, the escape energy loss timescale, 

\begin{equation}
\tau_{\rm esc}^{-1}=\tau_{\rm dif}^{-1}+\tau_{\rm adv}^{-1},
\end{equation}

\noindent may be energy-dependent since the diffusion time,
$\tau_{\rm dif}$, depends upon particle rigidity in interstellar magnetic
fields. For the Milky Way, the main contribution to the gamma-ray
luminosity comes from photons with energies $\sim$1 GeV,
corresponding to CR proton energies of $\sim$10 GeV. The diffusive
escape timescale can be estimated as $\tau_{\rm dif} \sim z_{h}^{2} / 6 D \sim (1-2) \times
10^7$ yr for a halo height of $z_{h}$ = 4 kpc, and taking diffusion
coefficients in the range $D = (4-8) \times 10^{28}$ cm$^{2}$ s$^{-1}$
for a particle rigidity of $\sim$10 GV
\citep{ptuskin_2006_diffusion}. Given a grammage of 12 g
cm$^{-3}$, the average gas density corresponding
to an escape time of $2 \times 10^7$ yr is $\langle n_{\rm
eff}\rangle \approx$ 0.3 cm$^{-3}$ (implying that $\tau_{\rm pp} \sim
1.5 \times 10^8$ yr). In spite of modeling differences, it is generally
understood that the diffusive escape timescale is much shorter than the collisional loss timescale
for the Milky Way \citep[][and references
therein]{strong_2007_cr_review}.


Large uncertainties exist for the escape time due to
uncertainty in diffusion parameters, and because CRs might also be advected
outwards by bulk motion `superwinds' which are observed for many
starbursts \citep{lehnert_1996_superwinds}. The advective timescale,
$\tau_{\rm adv}$, like the collisional energy loss timescale, is nearly independent of
energy.


The spectrum of pionic gamma rays will be affected by a transition
between dominant energy loss mechanisms, becoming softer
for galaxies in which energy-dependent diffusive losses are
important. In particular, the harder gamma-ray spectra found for
starburst galaxies may be understood if the energy
loss rate due to either proton-proton interactions or advection is considerably faster
than the diffusion timescale. A scaling relation for gamma-ray luminosity
should include the possibility that residence times for CR nuclei
in different galaxies may vary substantially.


In the calorimeteric limit, the residence time and effective density
of target material are inversely proportional, and accordingly, the anticipated scaling of
pionic gamma-ray luminosity becomes a linear scaling of the SFR:


\begin{align}
L_{0.1-100 \; {\rm GeV},\pi^{0}}\vert_{{\tau_{\rm
res}} \approx \tau_{\rm pp}} = 5 \times 10^{39} \rm{erg\;s}^{-1}
\left(\frac{\rm SFR}{{\rm M}_{\odot}{\rm yr}^{-1}}\right) \notag\\
\left(\frac{\mathit{E}_{\rm SN}}{10^{51} \rm{erg}}\right)
\left(\frac{\eta}{0.1}\right).
\label{eq_luminosity_calorimeter}
\end{align}

\noindent For the calorimeter case, we assume an underlying spectrum of CR nuclei described
by a power law with $\Gamma_{\rm p}=2.2$ ($\mathcal{Q}_{0.1-100 \; {\rm GeV}} = 7.8 \times
10^{-17}$ s$^{-1}$ H-atom$^{-1}$ cm$^3$), representative of
the LAT-detected starbursts. The gamma-ray luminosity expected in the
calorimetric limit for CR nuclei is plotted in Figures
\ref{fig_rc_correlation} and \ref{fig_ir_correlation} assuming an average CR
luminosity per supernova of $E_{\rm SN}\;\eta=10^{50}$
erg.\footnote{\cite{lacki_2011_starbursts_interpretation} calculated
the ratio $L_{>1 \; {\rm GeV}} / L_{8-1000 \; \mu{\rm m}} = 3.1 \times
10^{-4}$ expected in the calorimetric limit for CR nuclei with
$\Gamma_{\rm p} = 2.0$ and $E_{\rm SN} \eta = 10^{51}$ erg. Substituting the corresponding gamma-ray
emissivity, $\mathcal{Q}_{>1 \; {\rm GeV}} = 1.2 \times
10^{-16}$ s$^{-1}$ H-atom$^{-1}$ cm$^3$, into equation
\ref{eq_luminosity_calorimeter} yields $L_{>1 \; {\rm GeV}} /
L_{8-1000 \; \mu{\rm m}} = 2.5 \times 10^{-4}$, an agreement to within $\sim$20\%.}


To determine how well the theoretical calorimetic limit above describes nearby
starburst galaxies, we can transform the
observed scaling relation between gamma-ray luminosity and total
IR luminosity to a scaling relation between gamma-ray luminosity and SFR using
equation \ref{eq_sfr_ir}. This produces

\begin{equation}
L_{0.1-100 \; {\rm GeV}}\vert_{\rm scaling\;relation} = N \left(\frac{\rm
SFR}{{\rm M}_{\odot}{\rm yr}^{-1}}\frac{1}{1.7 \epsilon}\right)^{\alpha},
\end{equation}

\noindent where $N = 1.85_{-0.31}^{+0.37} \times 10^{39}$ erg s$^{-1}$ and $\alpha =
1.16\pm0.07$, fit using the EM algorithm for the complete sample of 69
galaxies. Substituting best-fit values for the normalization and
index, and assuming a Chabrier IMF ($\epsilon$ = 0.79) yields


\begin{align}
L_{0.1-100 \; {\rm GeV}}\vert_{\rm scaling\;relation} \approx (1.3 \pm 0.3) \times 10^{39} {\rm
erg\;s}^{-1} \notag\\
\left(\frac{\rm SFR}{{\rm M}_{\odot}{\rm yr}^{-1}}\right)^{1.16 \pm 0.07},
\label{eq_luminosity_sfr}
\end{align}

The ratio between the gamma-ray luminosity estimated via the observed
scaling relation with total IR luminosity (equation
\ref{eq_luminosity_sfr}) and predicted in the
calorimetric limit for CR nuclei (equation
\ref{eq_luminosity_calorimeter}) provides an estimate for the calorimetric
efficiency of CR nuclei:


\begin{align}
F_{\rm cal} \lesssim \frac{L_{0.1-100 \; {\rm GeV}}\vert_{\rm scaling\;relation}}{L_{0.1-100 \; {\rm GeV},\pi^{0}}\vert_{\tau_{\rm res}\approx\tau_{\rm pp}}} 
\approx (0.3 \pm 0.1) \notag\\
\left(\frac{\rm SFR}{{\rm M}_{\odot}{\rm yr}^{-1}}\right)^{0.16 \pm 0.07}
\left(\frac{E_{\rm SN}}{10^{51} \rm{erg}}\right)^{-1}
\left(\frac{\eta}{0.1}\right)^{-1}.
\label{eq_calorimetric_efficiency}
\end{align}

\noindent This calorimetric efficiency estimate should be viewed as an upper limit because
leptonic processes and individual sources must also contribute to the
observed global gamma-ray emission of galaxies. We may infer from equation
\ref{eq_calorimetric_efficiency} that starburst galaxies
with SFR $\sim 10 \; {\rm M}_{\odot}{\rm yr}^{-1}$ have maximum calorimetric
efficiencies of 30--50\% whereas dwarf galaxies with SFR $\sim 0.1
{\rm M}_{\odot}{\rm yr}^{-1}$ have lower calorimetric
efficiencies of 10--20\%, when assuming an average CR
luminosity per supernova of $E_{\rm SN}\;\eta=10^{50}$ erg, using a
Chabrier IMF, and attributing all observed gamma-rays to
hadronic processes. Real galaxies are expected to exhibit some dispersion about
the trend. The estimates presented here are in good
agreement with the calculations of
\cite{lacki_2011_starbursts_interpretation} for M82 and NGC 253. Our
own Milky Way is the only galaxy for which global gamma-ray luminosity can be compared to
direct CR measurements, and the CR energetics are in fair agreement
with our estimates above considering that the estimated gamma-ray luminosity
is below (but consistent with) the best-fit scaling relation.

Energy-independent cooling mechanisms for CR nuclei capable of preserving
the injected spectral shape are preferred for the LAT-detected
starburst galaxies given their relatively hard gamma-ray
spectra. An enhancement of interaction efficiency is suggested by the
slightly non-linear scaling relations found in Section
\ref{subsec_multiwavelength}. However, the overall normalizations of
these scaling relations imply that most starburst galaxies are not
fully calorimetric for $E_{\rm SN}\;\eta=10^{50}$ erg. Our present
analysis does not tightly constrain the CR calorimetry scenario for
ULIRGs. Energy-independent CR transport should also be examined
as a potentially important energy-loss mechanism for CR nuclei in starburst galaxies
\citep[see in this context][]{persic_2012_cr_energy_density}. Our
observations suggest that a substantial fraction of CR energy escapes into intergalactic space for
most galaxies since CR nuclei are expected to dominate the CR energy
budget, but that some starburst systems such as M82, NGC 253, NGC
1068, and NGC 4945 have substantially higher calorimetic
efficiencies relative to the Milky Way. A detailed discussion of
particular starburst galaxies as potential
CR calorimeters is presented by \cite{lacki_2011_starbursts_interpretation}.

%% file: subsec_egb.tex
\subsection{Contribution to the Isotropic Diffuse Gamma-ray Background}\label{subsec:egb}


Celestial gamma-ray radiation not resolved into individual
sources is often treated as the sum two components: a spatially structured
component originating mainly from CR interactions in the Milky Way, and a
nearly uniform all-sky component often called the `isotropic diffuse
gamma-ray background' (IGRB). Shortly after the discovery of an
isotropic component using the OSO-3 satellite \citep{clark_1968_oso_3} and early
spectral characterization with SAS-2 \citep{fichtel_1975_sas_2}, several
authors proposed that multitudinous extragalactic sources too faint to
be individually resolved must contribute to the observed flux
\citep{strong_1976_egb_origin,lichti_1978_galaxies_egb}. The spectrum
of the isotropic component has been subsequently measured with EGRET
\citep{sreekumar_1998_egret_egb,strong_2004_egret_egb}, and most
recently with the \textit{Fermi} LAT \citep{lat_2010_egb}. Note that the
intensity attributed to the isotropic diffuse component is \textit{instrument-dependent} in the sense
that more sensitive instruments are capable of extracting fainter
individual sources. In this work, the term IGRB will refer specifically to
the most recent measurement reported by the \textit{Fermi}-LAT
Collaboration, consistent with a featureless power law of spectral
index of $2.41\pm0.05$ between 0.2--100 GeV with integral intensity
above 0.1 GeV of $1.03\pm0.17 \times 10^{-5}$ ph cm$^{-2}$ s$^{-1}$ sr$^{-1}$.

The origin of the IGRB flux is not yet fully understood, in part
because the contribution from the most
prominent extragalactic gamma-ray source class, namely blazars, is
well-constrained by population synthesis techniques. More than one thousand sources
at high Galactic latitudes, predominantly blazars, have
now been discovered at GeV energies \citep{lat_2011_2lac}. The total
intensity of these resolved sources averaged over the full sky is $0.44 \times 10^{-5}$ ph cm$^{-2}$
s$^{-1}$ sr$^{-1}$. Based on the
empirically-determined flux distribution of high-latitude sources,
blazars with fluxes below the LAT detection threshold are expected to
contribute less than 30\% of the IGRB intensity \citep{lat_2010_high_latitude}. 


Star-forming galaxies far outnumber AGN in number density, but are
more challenging to detect in high-energy gamma rays due to their comparatively modest luminosities and lack of
beamed emission. Even with the improved sensitivities of contemporary gamma-ray telescopes, a limited
number of exclusively low-redshift star-forming galaxies are
expected to be individually detected. Estimates for the collective intensity of
unresolved galaxies consequently rely upon scaling relations between gamma-ray
luminosity and galaxy proporties (such as SFR and gas content), and studies of the evolving cosmological
population of galaxies. Calculations by \cite{pavlidou_2002_guaranteed} and \cite{thompson_2007_calorimeter}
during the EGRET era, and by \cite{fields_2010_galaxies_egb},
\cite{makiya_2011_galaxies_egb}, and \cite{stecker_2011_egb_components} incorporating
additional contraints from the first year of the \textit{Fermi}
mission demonstrated that star-forming galaxies could make a substantial contribution to the
IGRB, comparable to that of the blazars. 



\cite{fields_2010_galaxies_egb} pointed out that uncertainties in the estimated contribution can arise from a
degeneracy between density and luminosity evolution of star-forming
galaxies over cosmic time. In other words, a change in either the
density of galaxies having a given SFR (density
evolution), or a change in the SFR of individual galaxies (luminosity
evolution) could account for variations in total SFR density with
redshift. If the relationship between gamma-ray
luminosity and SFR is non-linear, such distinctions are important.

\cite{stecker_2011_egb_components} proposed an approach combining
global gamma-ray luminosity scaling relations with
multiwavelength luminosity functions which describe the abundances of
individual galaxies within different luminosity classes over
redshift. As an example of this approach, we now estimate the
contribution of non-AGN dominated star-forming
galaxies to the IGRB using the relationship between gamma-ray
luminosity (0.1--100 GeV) and total IR (8--1000 $\mu$m) luminosity
identified in section \ref{subsec_multiwavelength}, together with an IR
luminosity function provided by \textit{Spitzer} observations of the
VIMOS VLT Deep Survey and GOODS fields \citep{rodighiero_2010_spitzer}.


This procedure requires an IR luminosity function for exclusively non-AGN dominated
star-forming galaxies. In obtaining luminosity functions from
\textit{Spitzer} data, Rodighiero et al. utilized a catalog of
spectral templates to simultaneously determine the
redshift and spectral classification of each galaxy in their
dataset. Objects classified as type-I AGN were subsequently removed
from the sample. Although some AGN likely remain, the fractional contamination of IR-bight obscurred
AGN is expected to be small compared to non-AGN galaxies
\citep{ballantyne_2007_agn_ir,petric_2010_agn_ir}, such that the final
IR luminosity functions accurately represent the cosmic star formation
history. The IR luminosity function for non-AGN galaxies published by Rodighiero et
al. (given in the rest-frame  of the galaxies) is consistent with
those found from other IR surveys \citep{sanders_2003_iras_rbgs,chapman_2005_ir,le_floch_2005_ir,huynh_2007_ir,magnelli_2009_ir}.


The collective intensity (ph cm$^{-2}$ s$^{-1}$ sr$^{-1}$ GeV$^{-1}$)
of unresolved star-forming galaxies at observed photon energy $E_0$ can be computed using the line-of-sight
integral,


\begin{equation}
I(E_0)=\int_0^{z_{\rm max}} \int_{L_{{\gamma},{\rm
min}}}^{L_{{\gamma},{\rm max}}} \Phi(L_{\gamma},z) \frac{d^2V}{dz d\Omega}
\frac{dN}{dE}\left(L_{\gamma},E_0(1+z)\right) dL_{\gamma}dz,
\end{equation}

\noindent where $\Phi(L_{\gamma},z)=d^2N/(dVdL_{\gamma})$ expresses the number
density of galaxies per unit luminosity interval, and $dN/dE\left(L_{\gamma},E_0(1+z)\right)$ is the differential
photon flux of an individual galaxy with integral gamma-ray luminosity
$L_{\gamma}$ at redshift $z$. The factor $d^2V/(dz d\Omega)$
represents the comoving volume element per unit redshift and unit solid angle.
We transform the integral over gamma-ray luminosity into an integral over IR luminosity
using the scaling relation equation \ref{eq_scaling_ir} and fitted
parameters from Table \ref{table_results}.


Given the limited number of star-forming galaxies detected in high
energy gamma rays, the spectral transition from quiescent to starburst galaxies has not
yet been mapped in detail. Accordingly, two spectral model choices for star-forming galaxies in the
gamma-ray energy band are used in our calculation. First, we adopt a power
law spectral model with photon index 2.2, characteristic of
the LAT-detected starbursts. The second spectral model is based on
a model of the global emission from the Milky Way
\citep{strong_2010_global} scaled to the appropriate corresponding IR luminosity. These two spectral
models should be viewed as bracketing the expected contribution since
multiple galaxy types with different gamma-ray spectral
characteristics contribute to the IGRB, e.g. dwarfs, quiescent spirals, and starbursts. 


Only the redshift range $0<z<2.5$ is considered in this work because IR
luminosity functions are not yet well constrained beyond
$z\sim2.5$. We explicitly assume that the relationship between
gamma-ray and IR luminosity found for galaxies at $z\leq0.05$ holds for galaxies at
higher redshifts. Although we are unable to validate this assertion
directly, the closely-related RC-IR correlation does not show signs
of evolution up to $z\sim2$ \citep{ivison_2010_herschel}, and possibly to
redshifts $z\gtrsim4$ \citep{sargent_2010_radio_ir}. Those observations are
indicative of a consistent relationship between star formation and CRs
in galaxies over the past 10 Gyr. 


We account for the attenuation of gamma rays by interactions with the extragalactic
background light using the model of \cite{franceschini_2008_ebl}. The
contribution of resulting cascade emission to the IGRB is expected to
be faint compared to the primary component \citep{makiya_2011_galaxies_egb}. 


Estimates for the collective intensities of unresolved star-forming galaxies,
including both quiescent galaxies and starbursts, in the energy range
above 0.1 GeV are shown in Figure \ref{fig_egb_contribution} relative to the
first-year \textit{Fermi} LAT IGRB measurement. Shaded regions denote
the range of statistical and systematic uncertainties. We use the
scaling relation between gamma-ray luminosity and total IR luminosity obtained using the EM algorithm
with the full sample of 69 galaxies, allowing for 1 standard deviation of variation in the fitted parameters, and include
log-normal intrinsic scatter of gamma-ray luminosities. An additional
30\% uncertainty is allocated for the normalizaion of the IR
luminosity function. Using the methodology described above with the scaled Milky Way
spectral model, the estimated integral photon intensity of unresolved
star-forming galaxies with redshifts $0<z<2.5$ above 0.1 GeV is
$1.2_{-0.6}^{+1.2} \times 10^{-6}$ ph cm$^{-2}$ s$^{-1}$ sr$^{-1}$. If the power law
model is assumed, the estimated integral photon intensity is $0.8_{-0.4}^{+0.7}
\times 10^{-6}$ ph cm$^{-2}$ s$^{-1}$ sr$^{-1}$. These estimates cover a
range of 4--23\% of the integral IGRB intensity above 0.1 GeV measured
with the LAT, $1.03\pm0.17 \times 10^{-5}$ ph cm$^{-2}$ s$^{-1}$
sr$^{-1}$ \citep{lat_2010_egb}. Differential intensities for the two
spectral models are presented in Table \ref{table_egb_intensity}. The
range of intensities predicted in this work is consistent with previous estimates for the
intensity of unresolved star-forming galaxies shown in Figure \ref{fig_egb_contribution} for
comparison.



The relative contributions of star-forming galaxies to the IGRB
according to their redshift and IR luminosity class are shown in
Figure \ref{fig_egb_fractional_contribution}. In both panels, fractional
contributions are normalized to the estimated total intensity of
star-forming galaxies with redshifts $0<z<2.5$.
These panels provide important checks to the reliability of our
total estimate given the uncertainty in the relationship between gamma-ray and IR
luminosity for all galaxy types (e.g. no ULIRGs are detected at GeV
energies), and our incomplete knowledge of the IR luminosity function
at intermediate redshifts. The black dashed curve indicates the total
IR luminosity above which \cite{rodighiero_2010_spitzer} report completeness
in the sample of galaxies used to derive their published IR luminosity
functions. The luminosity
function in the region of phase space with IR luminosity
above that threshold is well-constrained by \textit{Spitzer} data. Although the IR luminosity
function for galaxies below this threshold is not directly determined
by the detections of individual galaxies, the total
density of IR emission is contrained by the observed extragalactic
background light. The IR luminositity functions presented by
\cite{rodighiero_2010_spitzer} integrated over IR luminosity and redshift
are consistent with the cosmic SFR history estimated from analyses of
IR observations by \cite{perez-gonzales_2005_ir},
\cite{le_floch_2005_ir}, and \cite{caputi_2007_ir}, and that estimated
with far-ultraviolet data \citep{tresse_2007_uv}. 

Although the contribution from star-forming galaxies is only
considered in the redshift range from $0<z<2.5$ in this work, Figure
\ref{fig_egb_fractional_contribution} makes apparent that the contribution
from star-forming galaxies at redshifts $z>1.5$ is diminishing. The
particular shape of the cumulative contribution curve in the lower
panel of Figure \ref{fig_egb_fractional_contribution} is determined mainly by the IR luminosity function.


Estimates for the combined intensities of unresolved blazars and star-forming
galaxies are plotted in Figure \ref{fig_egb_contribution_sum}. The
summed contribution falls short of explaining the IGRB, suggesting that important aspects of these populations are
not included in current models and/or that other high-energy source classes or
diffuse processes contribute a significant fraction of the observed
IGRB intensity. \cite{dermer_2007_egb} provides an extensive discussion of possible contributions to the IGRB.


Besides the beamed emission from blazars, in which the relativistic jet direction coincides with our
line of sight, the gamma-ray emission from the cores of misaligned AGN
must also constitute part of the IGRB. More than 10 such
radio galaxies have now been detected in \textit{Fermi} LAT data at
redshifts up to $z\sim0.7$ \citep{lat_2010_misaligned}. A relationship between the gamma-ray and radio emission
of LAT-detected misaligned AGN has been used to estimate that gamma-ray-loud radio
galaxies account for $\sim25\%$ of the unresolved IGRB above 0.1 GeV
\citep{inoue_2011_radio_galaxies}. However, estimates for this contribution are not yet
tightly constrained due to the small sample of LAT-detected objects and uncertainties
regarding the scaling relation between radio and gamma-ray
luminosities.


Large scale structure formation shocks leading to the assembly of
galaxy clusters represent another interesting candidate population to
explain the remaining IGRB intensity
\citep{colafrancesco_1998_clusters}. The tenuous intergalactic medium
of galaxy clusters is thought to be a reservoir for CR nuclei which accumulate over cosmological
timescales. Although non-thermal leptonic populations are
well-established by observation of Mpc-scale radio features, no galaxy cluster has been
detected in the GeV energy range \citep[e.g.][]{lat_2010_cluster}. Other extragalactic
source classes have been considered, such as gamma-ray bursts \citep{le_2007_grb}, and extended emission from the lobes of
radio galaxies \citep{stawarz_2006_fr1,massaro_2011_lobes}, but
these contributions are exprected to be less than 1\% in the GeV
energy range. Unresolved radio-quiet AGN are also not expected to make
a significant contribution to the IGRB \citep{teng_2011_seyfert}.


An example of a truly diffuse component is the cumulative emission
resulting from electromagnetic cascades of ultra-high energy CRs
interacting with cosmic microwave background photons
\citep{berezinskii_1975_neutrinos} and very-high energy photons interacting
with the extragalactic background light \citep{coppi_1997_cascade}. The gamma rays produced by those cascades are expected to have a
relatively hard spectrum. Already, the LAT IGRB measurement has been used
to constrain this component and to predict the cosmogenic ultra-high
energy neutrino flux originating from charged pion decays of
the ultra-high energy CR interactions \citep{ahlers_2010_gzk,berezinsky_2011_uhecr,wang_2011_uhecr}.


Galactic sources, such as a population of unresolved millisecond pulsars at high Galactic
latitudes, could become confused with isotropic diffuse emission as argued by
\cite{faucher-giguere_2010_msp}. Part of
the IGRB may also come from our Solar System as a result of CR interactions
with debris of the Oort Cloud \citep{moskalenko_2009_oort}. 


Finally, a portion of the IGRB may originate from `new physics' processes
involving, for instance, the annihilation or decay of dark matter
particles \citep{bergstrom_2001_dark_matter_egb,ullio_2002_dark_matter_egb,taylor_2003_dark_matter_egb}.


Studies of anisotropies in the IGRB intensity on small angular scales
provide another approach to identify IGRB constituent source
populations \citep{siegal-gaskins_2008_egb_anisotropy}. The fluctuation
angular power contributed by unresolved star-forming galaxies is expected to be small compared to other source
classes because star-forming galaxies have the highest spatial density
among confirmed extragalactic gamma-ray emitters, but are individually faint
\citep{ando_2009_galaxy_clustering_egb}. Unresolved star-forming
galaxies could in principle explain the entire IGRB intensity without exceeding the measured
anisotropy \citep{lat_2012_egb_anisotropy}. By contrast, the
fractional contributions of unresolved blazars and millisecond pulsars
to the IGRB intensity are constrained to be less than $\sim20$\% and
$\sim2$\%, respectively, due to larger angular power expected for those source classes.

%% file: subsec_predictions.tex
\section{Galaxy Detection Outlook for the \textit{Fermi} LAT}\label{subsec_predictions}

The scaling relations obtained in Section \ref{subsec_multiwavelength} allow straightforward
predictions for the next star-forming galaxies which could be detected
by the LAT. We use the relationship between gamma-ray luminosity and
total IR luminosity to select the most promising targets over a 10-year \textit{Fermi} mission.

We begin by creating an IR flux-limited sample of galaxies from the IRAS
Revised Bright Galaxies sample \citep{sanders_2003_iras_rbgs} by
selecting all the galaxies with 60 $\mu$m
flux density greater than 10 Jy (248 galaxies). Next, 0.1--100 GeV
gamma-ray fluxes of the galaxies are estimated using the scaling relation between
gamma-ray luminosity and total IR luminosity. Intrinsic dispersion in
the scaling relation is addressed by creating a distribution of
predicted gamma-ray fluxes for each galaxy, assuming the log-normal
intrinsic scatter fitted with the EM algorithm for the full sample of
examined galaxies. Finally, we estimate the LAT flux sensitivity at
the location of each galaxy given the Galactic foreground and
isotropic diffuse background, and extrapolating the current LAT observation
profile. For each realization of the population, we count galaxies
with predicted fluxes above the LAT sensitivity thresholds at their
respective positions as `detected.' The galaxies are modeled as spatially unresolved sources
having power law spectral forms with photon index equal to 2.2. These
modeling choices are focused on the search for additional starburst galaxies beyond the
Local Group.


Detection probabilities over a 10-year \textit{Fermi} mission are
plotted in Figure \ref{fig_detection_prob_src} for the best candidate
galaxies according to the procedure outlined above. The predictions
pass the consistency check that the best candidate galaxies match
those which have been LAT-detected. Galaxies with the next highest
probabilities of LAT-detection in the coming years include M33, M83,
NGC 3690, NGC 2146, and Arp 220. This list substantially overlaps with the
top candidates recently named by \cite{lacki_2011_starbursts_interpretation}.


Formally, the galaxy NGC 5128 (Centaurus A) would have entered our
list of top condidates based on its total IR flux, and indeed, NGC
5128 has been detected at both GeV and TeV energies
\citep{aharonian_2009_cen_a,lat_2010_cen_a}. However, the signal is thought to
originate mainly from the relativistic jet associated with the
AGN. NGC 5128 is consequently excluded from our list of candidate galaxies.

We find moderately significant excesses above backgrounds at the locations of NGC 2146
and M83 (see Section \ref{sec_results}). If those excesses are indeed
associated with the galaxies (see in this context
\cite{lenain_2011_local_group} for M83), then both might be securely detected after $\sim6$ years of LAT observations.


Figure \ref{fig_detection_prob_total} shows the cumulative number of
galaxies anticipated to be detected as a function of increasing mission time. The shaded
confidence regions are obtained directly from the distribution of
total galaxies above the LAT sensitivity threshold as predicted from
the multiple realizations of the galaxy population described
above. The actual numbers of external galaxies
reported in the 1FGL \citep{lat_2010_1fgl} and 2FGL
\citep{lat_2012_2fgl} catalogs are consistent with expectations. If
the simple scaling relationship between gamma-ray luminosity and
total IR luminosity found for our present sample is representative of
the larger population of low-redshift star-forming galaxies, then we
can expect about 10 external galaxies to be gamma-ray detected during a
10-year \textit{Fermi} mission.

%% file: sec_conclusions.tex
\section{Conclusions}\label{sec:conclusions} 

We examined 64 galaxies selected for their abundant dense molecular
gas using three years of \textit{Fermi} LAT data in the 0.1--100 GeV energy
range. Those results are combined with previous studies of 5 Local Group
galaxies in a multiwavelength analysis incorporating both measured
fluxes for the LAT-detected galaxies and gamma-ray flux upper limits. We find
further, though not yet conclusive ($P$-values $\sim0.05$ in the most
conservative case), evidence for a quasilinear scaling relation between
gamma-ray luminosity and star-formation rate as estimated by radio
continuum luminosity or total IR luminosity. Since contemporaneous star-formation rates are
sensitive to short-lived massive stars, and because gamma rays are the most
direct probe of CR hadrons in the ISM of external galaxies, the
scaling relationship strengthens the connection between CR energy content and massive
stars which ultimately explode as supernovae.


In the paradigm that SNRs channel approximately 10\% of their
mechanical energy into CR nuclei with kinetic energies $>1$ GeV, the
normalization of the observed scaling relationship between gamma-ray
luminosity and SFR implies that starburst galaxies (SFR $\sim10 \;
M_{\odot}$ yr$^{-1}$) have an average calorimetric efficiency for CR
nuclei of 30--50\% if the gamma-ray emission is
dominated by neutral pion decay. As discussed in Section
\ref{subsec_cr}, this constraint depends upon the assumed energetics of SNRs
and the stellar initial mass function. Meanwhile, the hard gamma-ray
spectra of starburst galaxies ($\Gamma$ = 2.2--2.3) give preference for
energy-independent loss mechanisms for CR nuclei in starbursts, as
opposed to the diffusive losses which likely shape the observed gamma-ray spectra of quiescent
Local Group galaxies. Both hadronic interactions and advective
transport of CR nuclei should be considered as potentially important energy loss
mechanisms in starburst galaxies. The overall normalization of the relation between gamma-ray
luminosity and SFR supports the understanding that the majority of CR energy in most galaxies
eventually escapes into intergalactic space.

The relationship between gamma-ray luminosity and total IR luminosity
can be used as a low-redshift anchor to estimate the contribution of
star-forming galaxies to the IGRB. Using the gamma-ray
luminosity scaling relations presented
here in conjunction with IR luminosity functions, we estimate that
star-forming galaxies with redshifts $0<z<2.5$ have a sky-averaged intensity of 0.4--2.4 $\times 10^{-6}$ ph cm$^{-2}$
s$^{-1}$ sr$^{-1}$ in the energy range 0.1--100 GeV, thereby
contributing 4--23\% of the IGRB intensity measured by the LAT. The combined contributions of unresolved blazars and
star-forming galaxies appear to fall short of fully explaining the observed
IGRB, suggesting that other gamma-ray source populations and/or truly
diffuse processes constitute a substantial fraction
of the observed IGRB intensity.

Finally, we predict that several more external galaxies might be
detected by the LAT during a 10-year \textit{Fermi} mission. In particular, the
galaxies M33, M83, NGC 3690, NGC 2146, and Arp 220 are interesting targets for further
study. The all-sky coverage provided by the LAT can also help identify promising
targets for studies with imaging air-Cherenkov
telescopes, including the proposed Cherenkov Telescope
Array\footnote{\url{http://www.cta-observatory.org/}}.

%% file: app_sec_kendall_tau.tex
\section{Kendall $\tau$ Correlation Statistic}\label{app_sec_kendall_tau}

The Kendall $\tau$ correlation statistic is a non-parametric
rank-correlation method which can be generalized for the analysis
of datasets containing both detections and non-detections. In this
work, we follow the procedure of \cite{akritas_siebert_1996_kendall} to test the
degree of correlation between luminosities in two wavebands for a collection of galaxies.

Consider a dataset consisting of $n$ points, $\mathbi{X}_k[i]$, indexed by $i=1,...,n$, with
$k={1,2}$ to denote the two wavebands. To compute the correlation
coefficient, first define a helper function,

\begin{equation}
J_k(i,j)=\delta(\mathbi{X}_k[i])I(\mathbi{X}_k[i]<\mathbi{X}_k[j])-\delta(\mathbi{X}_k[j])I(\mathbi{X}_k[j]<\mathbi{X}_k[i])),
\end{equation}

\noindent where $I=1$ if the enclosed conditional statment is true, and $I=0$
otherwise. When performing an analysis of partially censored data, let
$\delta(\mathbi{X}_k[i])=1$ for a detection (measured value), and $\delta(\mathbi{X}_k[i])=0$ for a
non-detection (limiting value). For historical reasons, the
Kendall $\tau$ statistic is conventionally defined in the context of
data containing lower limits rather than upper limits as is common in astrophysics
research applications. The method can be applied to data containing
upper limits by taking the negative value of all quantities,
i.e. $\mathbi{X}^{\prime}_k[i]=-\mathbi{X}_k[i]$. Next, define a
function to encode the ordering between two points in the dataset labeled $i,j$,

\begin{equation}
H(i,j)=J_1(i,j)J_2(i,j).
\end{equation}

A graphical representation of this encoding is shown in Figure \ref{fig_kendall_demo} for a dataset containing upper
limits in the $k=2$ waveband. Notice that $H(i,j)\in{-1,0,1}$. The Kendall $\tau$ correlation coefficient is the
sum of $H(i,j)$ for each pair of points in the dataset, normalized by
the total number of pairs,

\begin{equation}
\tau=\frac{2}{n(n-1)}\sum_i^n\sum_{j>i}^n H(i,j).
\end{equation}

\section{Potential Hazard of Multiwavelength Comparisons in Flux-Space}\label{app_sec_flux_hazard}

Consider the simple case of a sample of objects observed in two
wavebands, denoted by $k=1,2$, which follow a power law
relation in intrinsic luminosity, $\log L_2 = \alpha \log L_1 +
\beta$. An algebraic conversion of this equation from luminosity-space
to flux-space yields, 

\begin{equation}
\log F_2 = \alpha \log F_1 + 2 (\alpha-1) \log d + \rm{constant},
\end{equation}

\noindent where $d$ is the distance to the distance to each
object in the sample, $L_k=4\pi d^2 F_k$. Notice that if the
relationship between intrinsic luminosities is non-linear,
$\alpha\neq1$, the distance to each galaxy enters as an additional term in
the flux-space relation. This situation can lead to unpredictable
behavior, since the objects in the sample are, in general, located at
different distances to the observer.

\section{Simulations of Multiwavelength Comparisons in Luminosity-Space}\label{app_sec_simulations}

We discuss results from a Monte Carlo study of multiwavelength
comparisons in luminosity-space in the context of
the present work. One concern of comparing luminosities is that both
quantities have a shared dependence on the measured distances to the
objects. Common selection effects in observational astrophysics, e.g. that
more luminous objects can be detected at greater distances, therefore
have the potential to induce spurious apparent luminosity correlations
between wavebands.

We performed a series of simulations of a population of objects observed in two
wavebands to investigate this potential source of bias. The two wavebands are differentiated in that the
`selection' band, here called the `$X$-band,' is used to select a sample
of objects from a larger population, and the objects are then observed
in a second `search' band, here called the `$Y$-band.' The bands are asymmetric in that
partial censoring is possible in the $Y$ band, whereas all objects
are detected in the $X$ band. The galaxies examined in this work fit
the above description as a flux-limited sample selected according to their IR, CO-line, and HCN-line
fluxes, and subsequently observed in the gamma-ray band.

We test various scenarios considering different intrinsic
relationships between $X$- and $Y$-band luminosities, sample selection
approaches, and detection efficiencies. Throughout the simulations, we assume that
distances to the objects and fluxes are measured without error. This
assumption implies completeness down to the detection threshold flux level in the $Y$ band.

For each scenario, repeat the following
procedure many times:

\begin{enumerate}

\item Create a population of objects with randomly distributed
distances from the observer and luminosities in the $X$ band. 

\item Assume a relationship between intrinsic luminosities in the $X$
and $Y$ bands, specifically a power law
form with intrinsic dispersion $\mathcal{D}$:

\begin{equation}
\log L_Y = \alpha \log L_X + \beta + \mathcal{D}.
\label{eq_luminosity}
\end{equation}

Intrinsic dispersion in the relationship between
luminosities in the $X$- and $Y$-bands is taken to be normally
distributed in log-space with standard deviation $\sigma_\mathcal{D}$,
i.e. $\mathcal{D}=\mathcal{N}(0,\sigma_\mathcal{D})$.

\item Select a subset of objects from the full population. The subset
will comprise the sample of objects observed in both wavebands and
considered in subsequent statistical analysis.

\item Objects are considered detected (non-detected) in the $Y$ band if their
$Y$-band flux is above (below) the $Y$-band detection threshold flux
level. For the objects which are not detected in the $Y$ band, set the flux
upper limit at the $Y$-band detection threshold flux level. Determine
corresponding luminosities (possibly limits) for each object in the sample.  

\item Compute the Kendall $\tau$ correlation coefficient for the
relationship between luminosities in the $X$ and $Y$ bands for the
subset of selected objects. 

\end{enumerate} 

For the particular examples which follow, we generate a population of
objects representative of low-redshift star-forming galaxies. The $X$-band
luminosity distribution is drawn from the total IR luminosity function of
Rodighiero et al. (2010) at redshift $z=0$, with a minimum IR
luminosity of $10^8 L_{\odot}$. The volume considered is a cube of 200
Mpc on a side. Objects are assumed to be uniformly distributed in
Euclidean three-dimensional space.  

We consider two sample selection approaches: volume-limited, and
flux-limited in the $X$ band. In either case, a sample consisting of 100 objects is
selected for each trial. The selected sample in each trial amounts to approximately $0.06\%$
of objects in the full simulated population.

The parameters of each simulation scenario include the intrinsic
power law scaling index between $X$- and $Y$-band luminosities ($\alpha$), the standard deviation of
the intrinsic dispersion in the luminosity-luminosity relationship
($\sigma$, dex), and the detection efficiency of sample objects in the $Y$ band. For each
scenario, we perform 1000 trials. The distributions of
Kendall $\tau$ correlation coefficients for the different scenarios
are summarized in Table \ref{table_volume_limited} for the case of volume-limited
sample selection, and in Table \ref{table_flux_limited} for the case of
flux-limited sample selection.

Volume-limited samples are generally considered to be unbiased,
provided that a large enough volume can be sensitively probed so as to
contain a distribution of objects representative of the full
population. The results contained in Table \ref{table_volume_limited} demonstrate that
in scenarios with no intrinsic relationship between luminosities
($\alpha=0$), the distribution of correlation coefficients is
consistent with being centered on zero ($\tau=0$ corresponds to no correlation).
The correlation test is not biased towards finding a non-existent
correlation in this example. By contrast, the distribution of correlation coefficients for
scenarios in which an intrinsic linear correlation
was considered ($\alpha=1$) consistently tend toward positive values,
thereby correctly identifying a positive correlation.

These conclusions do not change appreciably in flux-limited sample
selection case, as evident in Table \ref{table_flux_limited}. For the simple
examples above, which are modelled on the population of
star-forming galaxies examined in this work in terms of luminosity
distribution, sample size, and sample selection, the Kendall
$\tau$ correlation coefficient proves to be effective in
distinguishing between populations with
and without intrinsically correlated luminosities, even when only a
small fraction of objects are actually detected in the `search' band.

We considered the simple case of power law relation between intrinsic
luminosities for this study. However, since the Kendall $\tau$
correlation test is a non-parametric method, the conclusions are
expected to hold for more complex scaling relations as well.

%% file: table_candidates.tex
\begin{table}\scriptsize 
\begin{center}
\caption{Summary of the star-forming galaxy sample: beyond the Local Group}
\label{table_candidates}
\begin{tabular}{c c c ccccc c c}
\\
\toprule
Galaxy & $D$ & $L_{1.4 \; \rm{GHz}}$ & $L_{12 \; \mu\rm{m}}$ &
$L_{25 \; \mu\rm{m}}$ & $L_{60 \; \mu\rm{m}}$ & $L_{100 \; \mu\rm{m}}$ &
$L_{8-1000 \; \mu\rm{m}}$ & $L_{\rm HCN}$ & \textit{Swift}-BAT Detected \\
 & (Mpc) & (10$^{21}$ W Hz$^{-1}$)  & \multicolumn{4}{c}{(10$^{23}$ W
  Hz$^{-1}$)} & (10$^{10}$ $L_{\odot}$) & (10$^{8}$ K km s$^{-1}$
pc$^{2}$) & (AGN Classification) \\
\midrule
NGC 253  & 2.5 & 4.18 & 0.307 & 1.16 & 7.24 & 9.63 & 2.1 & 0.27 & \nodata \\
M82  & 3.4 & 10.6 & 1.10 & 4.60 & 20.5 & 19.0 & 4.6 & 0.30 & \nodata \\
IC 342  & 3.7 & 3.69 & 0.244 & 0.565 & 2.96 & 6.42 & 1.4 & 0.47 & \nodata \\
NGC 4945  & 3.7 & 10.8 & 0.454 & 0.694 & 10.2 & 21.8 & 2.6 & 0.27 & Sy2 \\
M83  & 3.7 & 4.26 & 0.352 & 0.714 & 4.35 & 8.58 & 1.4 & 0.35 & \nodata \\
NGC 4826  & 4.7 & 0.268 & 0.0624 & 0.0756 & 0.970 & 2.16 & 0.26 & 0.040 & \nodata \\
NGC 6946  & 5.5 & 5.05 & 0.438 & 0.749 & 4.70 & 10.5 & 1.6 & 0.49 & \nodata \\
NGC 2903  & 6.2 & 2.06 & 0.243 & 0.397 & 2.78 & 6.00 & 0.83 & 0.090 & \nodata \\
NGC 5055  & 7.3 & 2.49 & 0.341 & 0.406 & 2.55 & 8.92 & 1.1 & 0.10 & \nodata \\
NGC 3628  & 7.6 & 3.63 & 0.216 & 0.335 & 3.79 & 7.31 & 1.0 & 0.24 & \nodata \\
NGC 3627  & 7.6 & 3.17 & 0.333 & 0.591 & 4.58 & 9.44 & 1.3 & 0.080 & \nodata \\
NGC 4631  & 8.1 & 9.42 & 0.405 & 0.704 & 6.70 & 12.6 & 2.0 & 0.080 & \nodata \\
NGC 4414  & 9.3 & 2.51 & 0.288 & 0.374 & 3.06 & 7.32 & 0.81 & 0.16 & \nodata \\
M51  & 9.6 & 16.4 & 0.795 & 1.05 & 10.7 & 24.4 & 4.2 & 0.50 & \nodata \\
NGC 891  & 10.3 & 8.90 & 0.669 & 0.889 & 8.44 & 21.9 & 2.6 & 0.25 & \nodata \\
NGC 3556  & 10.6 & 4.11 & 0.308 & 0.563 & 4.38 & 10.3 & 1.4 & 0.090 & \nodata \\
NGC 3893  & 13.9 & 3.30 & 0.335 & 0.381 & 3.60 & 8.51 & 1.2 & 0.23 & \nodata \\
NGC 660  & 14.0 & 9.08 & 0.715 & 1.71 & 15.4 & 26.9 & 3.7 & 0.26 & \nodata \\
NGC 5005  & 14.0 & 4.29 & 0.387 & 0.530 & 5.20 & 14.9 & 1.4 & 0.41 & \nodata \\
NGC 1055  & 14.8 & 5.58 & 0.587 & 0.744 & 6.12 & 17.1 & 2.1 & 0.37 & \nodata \\
NGC 7331  & 15.0 & 5.86 & 1.06 & 1.59 & 12.1 & 29.7 & 3.5 & 0.44 & \nodata \\
NGC 2146  & 15.2 & 29.7 & 1.89 & 5.20 & 40.6 & 53.6 & 10 & 0.96 & \nodata \\
NGC 3079  & 16.2 & 26.7 & 0.798 & 1.13 & 15.9 & 32.9 & 4.3 & 1.0 & Sy2 \\
NGC 1068  & 16.7 & 167 & 13.3 & 29.2 & 65.5 & 85.9 & 28 & 3.6 & Sy2 \\
NGC 4030  & 17.1 & 5.50 & 0.472 & 0.805 & 6.47 & 17.8 & 2.1 & 0.54 & \nodata \\
NGC 4041  & 18.0 & 4.04 & 0.438 & 0.605 & 5.49 & 12.3 & 1.7 & 0.18 & \nodata \\
NGC 1365  & 20.8 & 27.4 & 2.65 & 7.39 & 48.8 & 85.8 & 13 & 3.1 & Sy1.8 \\
NGC 1022  & 21.1 & 2.61 & 0.378 & 1.75 & 10.5 & 14.6 & 2.6 & 0.20 & \nodata \\
NGC 5775  & 21.3 & 15.2 & 0.993 & 1.34 & 12.8 & 30.2 & 3.8 & 0.57 & \nodata \\
NGC 5713  & 24.0 & 11.0 & 1.01 & 1.96 & 15.2 & 25.7 & 4.2 & 0.22 & \nodata \\
NGC 5678  & 27.8 & 10.3 & 0.869 & 1.11 & 8.94 & 23.7 & 3.0 & 0.75 & \nodata \\
NGC 520  & 31.1 & 20.5 & 1.04 & 3.73 & 36.5 & 54.8 & 8.5 & 0.64 & \nodata \\
NGC 7479  & 35.2 & 15.1 & 2.03 & 5.72 & 22.1 & 39.6 & 7.4 & 1.1 & Sy2/LINER \\
NGC 1530  & 35.4 & 10.4 & 1.08 & 1.84 & 14.8 & 38.7 & 4.7 & 0.49 & \nodata \\
NGC 2276  & 35.5 & 40.7 & 1.61 & 2.46 & 21.5 & 43.7 & 6.2 & 0.40 & \nodata \\
NGC 3147  & 39.5 & 17.3 & 3.64 & 1.92 & 15.3 & 55.3 & 6.2 & 0.90 & \nodata \\
Arp 299  & 43.0 & 150 & 8.78 & 54.2 & 250 & 246 & 63 & 2.1 & \nodata \\
IC 5179  & 46.2 & 43.4 & 3.01 & 6.13 & 49.5 & 95.2 & 14 & 3.4 & \nodata \\
NGC 5135  & 51.7 & 64.2 & 2.01 & 7.61 & 53.9 & 99.0 & 14 & 2.7 & \nodata \\
NGC 6701  & 56.8 & 35.6 & 2.12 & 5.10 & 38.8 & 77.4 & 11 & 1.4 & \nodata \\
NGC 7771  & 60.4 & 62.3 & 4.32 & 9.47 & 85.9 & 175 & 21 & 6.5 & \nodata \\
NGC 1614  & 63.2 & 66.1 & 6.60 & 35.8 & 154 & 164 & 39 & 1.3 & \nodata \\
NGC 7130  & 65.0 & 96.4 & 2.93 & 10.9 & 84.5 & 131 & 21 & 3.3 & Sy2/LINER \\
NGC 7469  & 67.5 & 98.7 & 8.67 & 32.5 & 149 & 192 & 41 & 2.2 & Sy1.2 \\
IRAS 18293-3413  & 72.1 & 141 & 7.09 & 24.8 & 222 & 332 & 54 & 4.0 & \nodata \\
Arp 220  & 74.7 & 218 & 4.07 & 53.4 & 695 & 770 & 140 & 9.2 & \nodata \\
Mrk 331  & 75.3 & 48.4 & 3.53 & 17.2 & 122 & 154 & 27 & 3.4 & \nodata \\
NGC 828  & 75.4 & 71.3 & 4.90 & 7.28 & 78.0 & 172 & 22 & 1.3 & \nodata \\
IC 1623  & 81.7 & 199 & 8.23 & 29.2 & 183 & 252 & 47 & 8.5 & \nodata \\
Arp 193  & 92.7 & 108 & 2.57 & 14.6 & 175 & 251 & 37 & 9.5 & \nodata \\
NGC 6240  & 98.1 & 492 & 6.79 & 40.9 & 264 & 305 & 61 & 11 & Sy2 \\
NGC 1144  & 117.3 & 256 & 4.58 & 10.4 & 87.3 & 187 & 25 & 2.7 & Sy2 \\
Mrk 1027  & 123.5 & 101 & 4.74 & 11.3 & 96.4 & 156 & 26 & 1.9 & \nodata \\
NGC 695  & 133.5 & 161 & 10.7 & 17.7 & 162 & 289 & 47 & 4.3 & \nodata \\
Arp 148  & 143.3 & 90.9 & 4.91 & 17.4 & 157 & 253 & 36 & 4.0 & \nodata \\
Mrk 273  & 152.2 & 403 & 6.65 & 65.4 & 624 & 624 & 130 & 15 & \nodata \\
UGC 05101  & 160.2 & 525 & 7.68 & 31.3 & 359 & 611 & 89 & 10 & \nodata \\
Arp 55  & 162.7 & 118 & 4.43 & 19.3 & 192 & 327 & 46 & 3.8 & \nodata \\
Mrk 231  & 170.3 & 1080 & 63.5 & 307 & 1070 & 1030 & 300 & 19 & \nodata \\
IRAS 05189-2524  & 170.3 & 102 & 25.7 & 120 & 460 & 411 & 120 & 6.2 & \nodata \\
IRAS 17208-0014  & 173.1 & 296 & 7.17 & 57.7 & 1150 & 1290 & 230 & 38 & \nodata \\
IRAS 10566+2448  & 173.3 & 208 & 7.19 & 45.6 & 435 & 539 & 94 & 10 & \nodata \\
VII Zw 31  & 223.4 & 245 & 11.9 & 37.0 & 329 & 603 & 87 & 9.8 & \nodata \\
IRAS 23365+3604  & 266.1 & 244 & 7.63 & 79.6 & 630 & 763 & 140 & 15 & \nodata \\
\bottomrule
\end{tabular}
\end{center}
\tablenotetext{}{Galaxy distances, total IR (8--1000 $\mu$m) luminosities, and
HCN line luminosities are provided by
\cite{gao_solomon_2004_dense_molecular_gas}. RC luminosities at 1.4 GHz come primarily
from \cite{yun_2001_radio}, except for M82 and NGC 3627 \citep{condon_1990_radio_bgs}, NGC 4945
\citep{wright_1990_parkes}, and Arp 299, NGC 5775, NGC 7331, and VII Zw
31 \citep{condon_1998_nvss}. Galaxies appearing in the \textit{Swift} BAT
58-month survey catalog with AGN-type classification are identified \citep{baumgartner_2010_swift}.}
\end{table}

%% file: table_local_group.tex
\begin{table}\scriptsize 
\begin{center}
\caption{Summary of star-forming galaxy sample: Local Group galaxies}
\label{table_local_group}
\begin{tabular}{c c c ccccc c c}
\\
\toprule
Galaxy & $D$ & $L_{1.4 \; \rm{GHz}}$ & $L_{12 \; \mu\rm{m}}$ & $L_{25
  \; \mu\rm{m}}$ & $L_{60 \; \mu\rm{m}}$ & $L_{100 \; \mu\rm{m}}$ & $L_{8-1000
  \; \mu\rm{m}}$ & $L_{\rm HCN}$ & $L_{0.1-100 \; \rm{GeV}}$ \\
 & (Mpc) & (10$^{20}$ W Hz$^{-1}$) & \multicolumn{4}{c}{(10$^{22}$ W Hz$^{-1}$)} & (10$^{9}$ $L_{\odot}$) & (10$^{7}$ K km s$^{-1}$ pc$^{2}$) & (10$^{38}$ erg s$^{-1}$) \\
\midrule
SMC & 0.06 & 0.19$\pm$0.03 & 0.0029 & 0.012 & 0.29 & 0.65 & 0.07$\pm$0.01 & \nodata & 0.11$\pm$0.03 \\
LMC & 0.05 & 1.3$\pm$0.1 & 0.083 & 0.23 & 2.5 & 2.5 & 0.7$\pm$0.1 & \nodata & 0.47$\pm$0.05 \\
M33 & 0.85 & 2.8$\pm$0.1 & 0.28 & 0.34 & 3.5 & 11 & 1.2$\pm$0.2 & \nodata & < 3.5 \\
M31 & 0.78 & 6.3$\pm$0.3 & 1.2 & 0.80 & 4.0 & 22 & 2.4$\pm$0.4 & \nodata & 4.6$\pm$1.0 \\
Milky Way & \nodata & 19$\pm$6 & 12 & 7.2 & 16 & 46 & 14$\pm$7 & 4$\pm$2 & 8.2$\pm$2.4 \\
\bottomrule
\end{tabular}
\end{center}
\tablenotetext{}{Global RC, IR, and gamma-ray luminosities of the Milky Way have been estimated using a numerical model of
CR propagation and interactions in the ISM \citep{strong_2010_global}. Within the Local Group, an
estimate for the global HCN line luminosity is only available for the Milky Way \citep{solomon_1992_dense_molecular_gas}. Radio data for other Local Group
galaxies: SMC \citep{loiseau_1987_smc_radio}, LMC \citep{hughes_2007_lmc_radio}, M31 and M33 \citep{dennison_1975_m31_m33_radio}.
IR data for other Local Group galaxies come from \cite{sanders_2003_iras_rbgs}.
Gamma-ray data for other Local Group galaxies: SMC \citep{lat_2010_smc}, LMC \citep{lat_2010_lmc}, M31 and M33
\citep{lat_2010_local_group}.}
\end{table}

%% file: table_results.tex
\begin{table}\scriptsize 
\begin{center}
\caption{Maximum likelihood analysis results}
\label{table_results}
\begin{tabular}{cc cccc ccc}
\\
\toprule
Galaxy & $D$ & $F_{0.1-100 \rm{GeV}}$ & $\Gamma$ & $L_{0.1-100 \rm{GeV}}$ & $TS$ \\
 & (Mpc) & (10$^{-9}$ ph cm$^{-2}$ s$^{-1}$) & & (10$^{40}$ erg
 s$^{-1}$) & \\
\midrule
NGC 253 & 2.5 & 12.6 $\pm$ 2.0 & 2.2 $\pm$ 0.1 & 0.6 $\pm$ 0.2 & 109.4 \\
M82 & 3.4 & 15.4 $\pm$ 1.9 & 2.2 $\pm$ 0.1 & 1.5 $\pm$ 0.3 & 180.1 \\
IC 342 & 3.7 & < 2.7 & 2.2 & 0.3 & \nodata \\
NGC 4945 & 3.7 & 8.5 $\pm$ 2.8 & 2.1 $\pm$ 0.2 & 1.2 $\pm$ 0.4 & 33.2 \\
M83 & 3.7 & < 7.0 & 2.2 & 0.8 & \nodata \\
NGC 4826 & 4.7 & < 2.9 & 2.2 & 0.6 & \nodata \\
NGC 6946 & 5.5 & < 1.6 & 2.2 & 0.4 & \nodata \\
NGC 2903 & 6.2 & < 2.5 & 2.2 & 0.8 & \nodata \\
NGC 5055 & 7.3 & < 2.1 & 2.2 & 1.0 & \nodata \\
NGC 3628 & 7.6 & < 3.1 & 2.2 & 1.6 & \nodata \\
NGC 3627 & 7.6 & < 3.5 & 2.2 & 1.7 & \nodata \\
NGC 4631 & 8.1 & < 1.6 & 2.2 & 0.9 & \nodata \\
NGC 4414 & 9.3 & < 3.1 & 2.2 & 2.3 & \nodata \\
M51 & 9.6 & < 3.2 & 2.2 & 2.6 & \nodata \\
NGC 891 & 10.3 & < 4.2 & 2.2 & 3.8 & \nodata \\
NGC 3556 & 10.6 & < 2.3 & 2.2 & 2.2 & \nodata \\
NGC 3893 & 13.9 & < 2.9 & 2.2 & 4.8 & \nodata \\
NGC 660 & 14.0 & < 3.0 & 2.2 & 5.0 & \nodata \\
NGC 5005 & 14.0 & < 3.4 & 2.2 & 5.7 & \nodata \\
NGC 1055 & 14.8 & < 2.9 & 2.2 & 5.5 & \nodata \\
NGC 7331 & 15.0 & < 1.7 & 2.2 & 3.2 & \nodata \\
NGC 2146 & 15.2 & < 6.7 & 2.2 & 13.2 & \nodata \\
NGC 3079 & 16.2 & < 2.2 & 2.2 & 5.0 & \nodata \\
NGC 1068 & 16.7 & 6.4 $\pm$ 2.0 & 2.2 $\pm$ 0.2 & 15.4 $\pm$ 6.1 & 38.1 \\
NGC 4030 & 17.1 & < 3.0 & 2.2 & 7.6 & \nodata \\
NGC 4041 & 18.0 & < 3.7 & 2.2 & 10.3 & \nodata \\
NGC 1365 & 20.8 & < 2.5 & 2.2 & 9.4 & \nodata \\
NGC 1022 & 21.1 & < 2.2 & 2.2 & 8.5 & \nodata \\
NGC 5775 & 21.3 & < 1.6 & 2.2 & 6.4 & \nodata \\
NGC 5713 & 24.0 & < 1.8 & 2.2 & 9.0 & \nodata \\
NGC 5678 & 27.8 & < 2.8 & 2.2 & 18.3 & \nodata \\
NGC 520 & 31.1 & < 2.0 & 2.2 & 16.7 & \nodata \\
NGC 7479 & 35.2 & < 6.1 & 2.2 & 65.2 & \nodata \\
NGC 1530 & 35.4 & < 2.8 & 2.2 & 29.7 & \nodata \\
NGC 2276 & 35.5 & < 1.4 & 2.2 & 15.5 & \nodata \\
NGC 3147 & 39.5 & < 1.8 & 2.2 & 23.5 & \nodata \\
Arp 299 & 43.0 & < 3.1 & 2.2 & 49.3 & \nodata \\
IC 5179 & 46.2 & < 0.9 & 2.2 & 17.3 & \nodata \\
NGC 5135 & 51.7 & < 1.5 & 2.2 & 34.2 & \nodata \\
NGC 6701 & 56.8 & < 2.4 & 2.2 & 65.2 & \nodata \\
NGC 7771 & 60.4 & < 2.0 & 2.2 & 62.4 & \nodata \\
NGC 1614 & 63.2 & < 2.1 & 2.2 & 72.2 & \nodata \\
NGC 7130 & 65.0 & < 1.3 & 2.2 & 47.5 & \nodata \\
NGC 7469 & 67.5 & < 2.4 & 2.2 & 94.9 & \nodata \\
IRAS 18293-3413 & 72.1 & < 2.0 & 2.2 & 90.8 & \nodata \\
Arp 220 & 74.7 & < 4.4 & 2.2 & 209.6 & \nodata \\
Mrk 331 & 75.3 & < 1.5 & 2.2 & 71.9 & \nodata \\
NGC 828 & 75.4 & < 2.7 & 2.2 & 129.9 & \nodata \\
IC 1623 & 81.7 & < 1.8 & 2.2 & 100.6 & \nodata \\
Arp 193 & 92.7 & < 2.6 & 2.2 & 189.2 & \nodata \\
NGC 6240 & 98.1 & < 2.3 & 2.2 & 186.5 & \nodata \\
NGC 1144 & 117.3 & < 1.2 & 2.2 & 141.1 & \nodata \\
Mrk 1027 & 123.5 & < 5.6 & 2.2 & 731.8 & \nodata \\
NGC 695 & 133.5 & < 4.2 & 2.2 & 646.1 & \nodata \\
Arp 148 & 143.3 & < 5.2 & 2.2 & 923.8 & \nodata \\
Mrk 273 & 152.2 & < 2.9 & 2.2 & 582.0 & \nodata \\
UGC 05101 & 160.2 & < 1.3 & 2.2 & 287.4 & \nodata \\
Arp 55 & 162.7 & < 1.7 & 2.2 & 380.1 & \nodata \\
Mrk 231 & 170.3 & < 1.9 & 2.2 & 468.4 & \nodata \\
IRAS 05189-2524 & 170.3 & < 1.6 & 2.2 & 395.6 & \nodata \\
IRAS 17208-0014 & 173.1 & < 5.6 & 2.2 & 1434.0 & \nodata \\
IRAS 10566+2448 & 173.3 & < 2.0 & 2.2 & 521.0 & \nodata \\
VII Zw 31 & 223.4 & < 1.6 & 2.2 & 692.6 & \nodata \\
IRAS 23365+3604 & 266.1 & < 1.7 & 2.2 & 1059.0 & \nodata \\
\bottomrule
\end{tabular}
\end{center}
\tablenotetext{}{Each galaxy is analyzed using a power law spectral model,
$dN/dE\propto E^{-\Gamma}$. Flux upper limits for galaxies not significantly detected in LAT
  data are presented at the 95\% confidence level, assuming a photon index $\Gamma$=2.2.}
\end{table}

%% file: table_significance.tex
\begin{table}\footnotesize
\begin{center}
\caption{Scaling relationships between global luminosities: non-parametric analysis of correlation significance}
\label{table_significance}
\begin{tabular}{c ccc ccc}
\\
\toprule
 & \multicolumn{3}{c}{Full Sample} & \multicolumn{3}{c}{Excluding AGN} \\
 & $\sigma_{D}$ = 0\% & $\sigma_{D}$ = 10\% & $\sigma_{D}$ = 20\% & $\sigma_{D}$ = 0\% & $\sigma_{D}$ = 10\% & $\sigma_{D}$ = 20\% \\
\midrule
$L_{1.4 \; {\rm GHz}}$ : $L_{0.1-100 \; {\rm GeV}}$ & 0.001 & 0.005 & 0.01 & 0.007 & 0.03 & 0.06 \\
$L_{8-1000 \; \mu {\rm m}}$ : $L_{0.1-100 \; {\rm GeV}}$ & 0.0003 & 0.001 & 0.005 & 0.0004 & 0.004 & 0.02 \\
$L_{\rm HCN}$ : $L_{0.1-100 {\rm GeV}}$ & 0.05 & 0.07 & 0.1 & 0.2 & 0.3 & 0.4 \\
\bottomrule
\end{tabular}
\end{center}
\tablenotetext{}{Results for the Kendall $\tau$ significance tests for
  correlation between gamma-ray luminosity and each of RC luminosity,
  IR luminosity, and HCN line luminosity, expressed as $P$-values representing the probabilities of
erroneously rejecting the null hypothesis that no correlation exists
 between wavebands. A description of the Kendall $\tau$ statistic is
provided in Appendix \ref{app_sec_kendall_tau} and details of the
implementation are provided in Section \ref{subsec_multiwavelength}. Scatter in the distance
measurements to the galaxies are assumed to be normally
distributed with standard deviation ($\sigma_{D}$) expressed as a
 percentage of the actual galaxy distances.}
\end{table}

%% file: table_regression.tex
\begin{table}\footnotesize
\begin{center}
\caption{Scaling relationships between global luminosities: power-law fits with full sample}
\label{table_regression}
\begin{tabular}{c ccc ccc}
\\
\toprule
 & \multicolumn{3}{c}{Expectation-Maximization Method} & \multicolumn{3}{c}{Buckley-James Method} \\
 & $\alpha$ & $\beta$ & $\sqrt{{\rm Variance}}$ & $\alpha$ & $\beta$ & $\sqrt{{\rm Variance}}$ \\
\midrule
\multicolumn{7}{c}{Full sample of galaxies} \\
\midrule
$L_{1.4 \; {\rm GHz}}$ : $L_{0.1-100 \; {\rm GeV}}$ & 1.10 $\pm$ 0.05 & 38.82 $\pm$ 0.06 & 0.17 & 1.10 $\pm$ 0.06 & 38.81 & 0.20 \\
(10$^{21}$ W Hz$^{-1}$) : (erg s$^{-1}$) & & & & & & \\
\midrule
$L_{8-1000 \; \mu {\rm m}}$ : $L_{0.1-100 \; {\rm GeV}}$ & 1.17 $\pm$ 0.07 & 39.28 $\pm$ 0.08 & 0.24 & 1.18 $\pm$ 0.10 & 39.31 & 0.31 \\
(10$^{10}$ $L_{\odot}$) : (erg s$^{-1}$) & & & & & & \\
\midrule
\multicolumn{7}{c}{Excluding galaxies hosting \textit{Swift}-BAT
  detected AGN} \\
\midrule
$L_{1.4 {\rm GHz}}$ : $L_{0.1-100 {\rm GeV}}$ & 1.10 $\pm$ 0.07 & 38.81 $\pm$ 0.07 & 0.19 & 1.09 $\pm$ 0.11 & 38.80 & 0.24 \\
(10$^{21}$ W Hz$^{-1}$) : (erg s$^{-1}$) & & & & & & \\
\midrule
$L_{8-1000 \mu {\rm m}}$ : $L_{0.1-100 {\rm GeV}}$ & 1.09 $\pm$ 0.10 & 39.19 $\pm$ 0.10 & 0.25 & 1.10 $\pm$ 0.14 & 39.22 & 0.33 \\
(10$^{10}$ $L_{\odot}$) : (erg s$^{-1}$) & & & & & & \\
\bottomrule
\end{tabular}
\end{center}
\tablenotetext{}{Fitted parameters for relationships between gamma-ray luminosity and multiwavelength tracers of
star-formation. Using the RC case as an example, the scaling relations are
of the form $\log L_{0.1-100 {\rm GeV}} = \alpha \log L_{1.4
{\rm GHz}} + \beta$, with luminosities expressed in the units provided
in the leftmost column. The square root of the variance provides an estimate of the intrinsic
 dispersion of gamma-ray luminosity residuals in log-space about the
 best-fit regression line. For the EM algorithm, the
 intrinsic residuals about the best-fit line are assumed to be
 normally distributed in log-space. The Buckley-James algorithm uses the
 Kaplan-Meier method to estimate the distribution of residuals. See
 Section \ref{subsec_multiwavelength} for a
 description of the fitting methods. Both the complete sample of 69
 galaxies (containing 8 LAT sources) and a subsample of 60 galaxies
 excluding the AGN detected by the \textit{Swift} BAT are analyzed (containing 6 LAT sources).}
\end{table}

%% file: table_egb_intensity.tex
\begin{table}\footnotesize
\begin{center}
\caption{Collective gamma-ray intensity of unresolved star-forming
  galaxies relative to the IGRB}
\label{table_egb_intensity}
\begin{tabular}{c c c c}
\\
\toprule
 & \multicolumn{2}{c}{Star-forming Galaxies} & Isotropic Diffuse \\
 & Re-scaled Milky Way Model & Power Law Model, $\Gamma=2.2$ & \\
E & $E^2 dN/dE$ & $E^2 dN/dE$ & $E^2 dN/dE$ \\
(GeV) & ($10^{-8}$ GeV cm$^{-2}$ s$^{-1}$ sr$^{-1}$) & ($10^{-8}$ GeV
cm$^{-2}$ s$^{-1}$ sr$^{-1}$) & ($10^{-8}$ GeV cm$^{-2}$ s$^{-1}$
sr$^{-1}$) \\
\midrule
0.10 & 4.0--15 & 4.9--18 & 150 \\
0.16 & 4.9--18 & 4.5--16 & 120 \\
0.25 & 5.4--20 & 4.1--15 & 100 \\
0.40 & 5.4--20 & 3.7--14 & 82 \\
0.63 & 4.9--18 & 3.4--12 & 68 \\
1.00 & 4.2--15 & 3.1--11 & 57 \\
1.6 & 3.3--12 & 2.8--10 & 47 \\
2.5 & 2.5--9.0 & 2.6--9.4 & 39 \\
4.0 & 1.9--6.7 & 2.4--8.6 & 32 \\
6.3 & 1.4--4.9 & 2.1--7.8 & 27 \\
10.0 & 1.0--3.6 & 1.9--7.1 & 22 \\
16 & 0.75--2.7 & 1.8--6.5 & 18 \\
25 & 0.55--2.0 & 1.6--5.8 & 15 \\
40 & 0.39--1.4 & 1.4--5.1 & 12 \\
63 & 0.26--0.94 & 1.2--4.3 & 10 \\
100 & 0.16--0.56 & 0.89--3.2 & 8.6 \\
160 & 0.082--0.28 & 0.59--2.0 & \nodata \\
250 & 0.036--0.12 & 0.34--1.1 & \nodata \\
400 & 0.022--0.078 & 0.19--0.63 & \nodata \\
\bottomrule
\end{tabular}
\end{center}
\tablenotetext{}{Estimated intensities of unresolved star-forming
  galaxies using two different spectral model assumptions
  are compared to the IGRB spectrum measured by with the
  LAT. The spectrum of the IGRB is consistent with a power law characterized by
  a photon index $\Gamma=2.41\pm0.05$ with an integral photon intensity
  above 0.1 GeV of $(1.03\pm0.17) \times 10^{-5}$ ph cm$^{-2}$
  s$^{-1}$ sr$^{-1}$ \citep{lat_2010_egb}. The entries in the
  rightmost column follow from this parametrization for the IGRB spectrum.}
\end{table}

%% file: table_volume_limited.tex
\begin{table}\footnotesize 
\begin{center}
\caption{Kendall $\tau$ Correlation Coefficient Distributions:
Volume-limited Sample}
\label{table_volume_limited}
\begin{tabular}{c cc}
\\
\toprule
\multicolumn{3}{c}{Detection Efficiency in $Y$-band = 0.1} \\
\midrule
 & $\alpha_{\rm intrinsic}=0$ & $\alpha_{\rm intrinsic}=1$ \\
\midrule
$\sigma_\mathcal{D}$=0 & 0 $\pm$ 0 & 0.12 $\pm$ 0.023 \\
 & (\nodata) & (5.1) \\
$\sigma_\mathcal{D}$=0.3 & -0.00021 $\pm$ 0.012 & 0.11 $\pm$ 0.022 \\
 & (-0.017) & (5.3) \\
$\sigma_\mathcal{D}$=1 & -0.002 $\pm$ 0.028 & 0.092 $\pm$ 0.024 \\
 & (-0.07) & (3.9) \\
\midrule
\multicolumn{3}{c}{Detection Efficiency in $Y$-band = 1} \\
\midrule
 & $\alpha_{\rm intrinsic}=0$ & $\alpha_{\rm intrinsic}=1$ \\
\midrule
$\sigma_\mathcal{D}$=0 & 0 $\pm$ 0 & 1 $\pm$ 0 \\
 & (\nodata) & (\nodata) \\
$\sigma_\mathcal{D}$=0.3 & -0.0018 $\pm$ 0.067 & 0.71 $\pm$ 0.03 \\
 & (-0.026) & (24) \\
$\sigma_\mathcal{D}$=1 & -0.0023 $\pm$ 0.068 & 0.36 $\pm$ 0.058 \\
 & (-0.035) & (6.2) \\
\bottomrule
\end{tabular}
\end{center}
\tablenotetext{}{Mean Kendall $\tau$ correlation coefficient
 and standard deviation of the distribution are supplied for each set
 of simulation parameters: detection efficiency in the $Y$ band, true
 power law scaling index of intrinsic luminosities ($\alpha$), and standard deviation of intrinsic
 dispersion in the luminosity-luminosity relationship ($\sigma_\mathcal{D}$,
 dex). 1000 realizations were analyzed for each set of simulation
 parameters. Below the mean and standard deviation of the distribution
 of correlation coefficients, and indicated in parentheses, is the
 ratio of the mean to the standard deviation.}
\end{table}

%% file: table_flux_limited.tex
\begin{table}\footnotesize 
\begin{center}
\caption{Kendall $\tau$ Correlation Coefficient Distributions:
 Flux-limited Sample}
\label{table_flux_limited}
\begin{tabular}{c cc}
\\
\toprule
\multicolumn{3}{c}{Detection Efficiency in $Y$-band = 0.1} \\
\midrule
 & $\alpha_{\rm intrinsic}=0$ & $\alpha_{\rm intrinsic}=1$ \\
\midrule
$\sigma_\mathcal{D}$=0 & 0 $\pm$ 0 & 0.068 $\pm$ 0.017 \\
 & (\nodata) & (4.1) \\
$\sigma_\mathcal{D}$=0.3 & 0.00074 $\pm$ 0.0056 & 0.072 $\pm$ 0.018 \\
 & (0.13) & (4.1) \\
$\sigma_\mathcal{D}$=1 & 0.0027 $\pm$ 0.017 & 0.065 $\pm$ 0.021 \\
 & (0.16) & (3.1) \\
\midrule
\multicolumn{3}{c}{Detection Efficiency in $Y$-band = 1} \\
\midrule
 & $\alpha_{\rm intrinsic}=0$ & $\alpha_{\rm intrinsic}=1$ \\
\midrule
$\sigma_\mathcal{D}$=0 & 0 $\pm$ 0 & 1 $\pm$ 0 \\
 & (\nodata) & (\nodata) \\
$\sigma_\mathcal{D}$=0.3 & 0.00004 $\pm$ 0.067 & 0.73 $\pm$ 0.03 \\
 & (0.00055) & (24) \\
$\sigma_\mathcal{D}$=1 & 0.0018 $\pm$ 0.07 & 0.38 $\pm$ 0.057 \\
 & (0.025) & (6.6) \\
\bottomrule
\end{tabular}
\end{center}
\tablenotetext{}{As Table \ref{table_volume_limited}, but considering
 flux-limited samples of objects in $X$-band flux.}
\end{table}